\documentstyle[preprint,eqsecnum,epsf,aps,floats]{revtex}
\tightenlines

\begin{document}
\title{Measurement of $W$ and $Z$ boson production cross sections}
%
%
\author{                                                                      
B.~Abbott,$^{40}$                                                             
M.~Abolins,$^{37}$                                                            
V.~Abramov,$^{15}$                                                            
B.S.~Acharya,$^{8}$                                                           
I.~Adam,$^{39}$                                                               
D.L.~Adams,$^{49}$                                                            
M.~Adams,$^{24}$                                                              
S.~Ahn,$^{23}$                                                                
G.A.~Alves,$^{2}$                                                             
N.~Amos,$^{36}$
% ...
\\                                                                            
\vskip 0.40cm                                                                 
\centerline{(D\O\ Collaboration)}                                             
\vskip 0.40cm                                                                 
}                                                                             
\address{                                                                     
\centerline{$^{1}$Universidad de Buenos Aires, Buenos Aires, Argentina}       
\centerline{$^{2}$LAFEX, Centro Brasileiro de Pesquisas F{\'\i}sicas,         
                  Rio de Janeiro, Brazil}                                     
\centerline{$^{3}$Universidade do Estado do Rio de Janeiro,                   
                  Rio de Janeiro, Brazil}                                     
\centerline{$^{4}$Universidad de los Andes, Bogot\'{a}, Colombia}             
\centerline{$^{5}$DAPNIA/Service de Physique des Particules, CEA, Saclay,     
                  France}                                                     
\centerline{$^{6}$Panjab University, Chandigarh, India}                       
\centerline{$^{7}$Delhi University, Delhi, India}                             
\centerline{$^{8}$Tata Institute of Fundamental Research, Mumbai, India}      
\centerline{$^{9}$Kyungsung University, Pusan, Korea}                         
\centerline{$^{10}$Seoul National University, Seoul, Korea}                   
\centerline{$^{11}$CINVESTAV, Mexico City, Mexico}                            
\centerline{$^{12}$Institute of Nuclear Physics, Krak\'ow, Poland}            
\centerline{$^{13}$Institute for Theoretical and Experimental Physics,        
                   Moscow, Russia}                                            
\centerline{$^{14}$Moscow State University, Moscow, Russia}                   
\centerline{$^{15}$Institute for High Energy Physics, Protvino, Russia}       
\centerline{$^{16}$University of Arizona, Tucson, Arizona 85721}              
\centerline{$^{17}$Lawrence Berkeley National Laboratory and University of    
                   California, Berkeley, California 94720}                    
\centerline{$^{18}$University of California, Davis, California 95616}         
\centerline{$^{19}$University of California, Irvine, California 92697}        
\centerline{$^{20}$University of California, Riverside, California 92521}     
\centerline{$^{21}$Florida State University, Tallahassee, Florida 32306}      
\centerline{$^{22}$University of Hawaii, Honolulu, Hawaii 96822}              
\centerline{$^{23}$Fermi National Accelerator Laboratory, Batavia,            
                   Illinois 60510}                                            
\centerline{$^{24}$University of Illinois at Chicago, Chicago,                
                   Illinois 60607}                                            
\centerline{$^{25}$Northern Illinois University, DeKalb, Illinois 60115}      
\centerline{$^{26}$Northwestern University, Evanston, Illinois 60208}         
\centerline{$^{27}$Indiana University, Bloomington, Indiana 47405}            
\centerline{$^{28}$University of Notre Dame, Notre Dame, Indiana 46556}       
\centerline{$^{29}$Purdue University, West Lafayette, Indiana 47907}          
\centerline{$^{30}$Iowa State University, Ames, Iowa 50011}                   
\centerline{$^{31}$University of Kansas, Lawrence, Kansas 66045}              
\centerline{$^{32}$Louisiana Tech University, Ruston, Louisiana 71272}        
\centerline{$^{33}$University of Maryland, College Park, Maryland 20742}      
\centerline{$^{34}$Boston University, Boston, Massachusetts 02215}            
\centerline{$^{35}$Northeastern University, Boston, Massachusetts 02115}      
\centerline{$^{36}$University of Michigan, Ann Arbor, Michigan 48109}         
\centerline{$^{37}$Michigan State University, East Lansing, Michigan 48824}   
\centerline{$^{38}$University of Nebraska, Lincoln, Nebraska 68588}           
\centerline{$^{39}$Columbia University, New York, New York 10027}             
\centerline{$^{40}$New York University, New York, New York 10003}             
\centerline{$^{41}$University of Rochester, Rochester, New York 14627}        
\centerline{$^{42}$State University of New York, Stony Brook,                 
                   New York 11794}                                            
\centerline{$^{43}$Brookhaven National Laboratory, Upton, New York 11973}     
\centerline{$^{44}$Langston University, Langston, Oklahoma 73050}             
\centerline{$^{45}$University of Oklahoma, Norman, Oklahoma 73019}            
\centerline{$^{46}$Brown University, Providence, Rhode Island 02912}          
\centerline{$^{47}$University of Texas, Arlington, Texas 76019}               
\centerline{$^{48}$Texas A\&M University, College Station, Texas 77843}       
\centerline{$^{49}$Rice University, Houston, Texas 77005}                     
}                                                                             

\maketitle

\begin{abstract}
D\O\ has measured the inclusive production cross section of $W$ and $Z$ bosons
in a sample of 13 pb$^{-1}$ of data collected at the Fermilab Tevatron.
The cross sections, multiplied by their leptonic branching fractions, 
for production in {\mbox{$p\bar p$}}\ collisions at 
{\mbox{$\sqrt{s}$ =\ 1.8\ TeV}}\
are
$\sigma_W 
{\rm B}({\mbox{$ W\rightarrow e \nu$}})=
2.36 \pm 0.02 \pm 0.08  \pm 0.13 {\rm~nb}$,
$\sigma_W 
{\rm B}({\mbox{$ W\rightarrow \mu \nu$}})=
2.09 \pm 0.06 \pm 0.22 \pm 0.11 {\rm~nb}$,
$\sigma_Z 
{\rm B}({\mbox{$ Z\rightarrow {e^+e^-}$}})=
0.218 \pm 0.008 \pm 0.008 \pm 0.012 {\rm~nb}$, and
$\sigma_Z 
{\rm B}({\mbox{$ Z\rightarrow {\mu^+\mu^-}$}})=
0.178 \pm 0.022 \pm 0.021 \pm 0.009 {\rm~nb}$,
where the first uncertainty is statistical and the second  systematic; 
the third reflects the uncertainty in the integrated luminosity.
For the combined electron and muon analyses, we find
${\sigma_W 
{\rm B}(W\rightarrow l\nu)}/
{\sigma_Z 
{\rm B}(Z\rightarrow l^+l^-)}=
10.90\pm 0.52$.
Assuming standard model couplings, we use this result  to determine 
the width of the $W$ boson, and obtain $\Gamma(W) = 2.044 \pm 0.097 \; 
{\rm GeV}$.
\end{abstract}
\section{Introduction}

  Measurement of the production cross sections multiplied 
by the leptonic branching fractions
($\sigma 
{\rm B}$) for $W$ and $Z$ bosons can be used to test 
predictions of QCD for $W$ and $Z$ production, and to extract 
the width of the $W$ boson.
Previous measurements of these cross sections
have been made at $\sqrt{s} = 630$ GeV by the UA1~\cite{UA1} and UA2~\cite{UA2}
experiments and at $\sqrt{s} = 1.8$ TeV by 
CDF~\cite{CDF1aX,CDF2aX,CDF3aX,CDF4aX,CDF5aX}.
The results reported in this paper
are from the D\O\ detector, operating at $\sqrt{s} = 1.8$ TeV,
and have been summarized previously in Ref.~\cite{D0PRL}.

   Precise determination of the total widths of the $W$ and $Z$ bosons
provides  an
important test of the standard model because these widths
are sensitive to new (and possibly undetected) decay modes.
The total width of the $Z$ boson is known to
a precision of $0.3\%$~\cite{PDG} which
places strong constraints on the existence of any new particles 
that can contribute to 
decays to neutrals.  Our knowledge of the 
total width of the $W$ boson is an order of 
magnitude less precise, and the
corresponding limits on charged weak decays are much less stringent.
It is therefore important to improve the measurement of the 
width of the $W$ boson 
as a means of searching for any unexpected $W$-boson decay modes.

We determine the width of the $W$ boson indirectly by using
the ratio of the measured $W$ and $Z$ boson $\sigma 
{\rm B}$ values
\begin{eqnarray*}
R &\equiv &
{{\sigma_W 
{\rm B}(W\rightarrow l\nu)} \over
 {\sigma_Z 
{\rm B}(Z\rightarrow ll)}},
\end{eqnarray*}
where $l$ corresponds to $e$ or $\mu$, 
$\sigma_W$ and $\sigma_Z$ are the inclusive
cross sections for $W$ and $Z$ boson production, 
$\sigma(p\bar p \to W + X)$ and $\sigma(p\bar p \to Z + X)$,
and ${\rm B}(W\rightarrow l\nu)$ and ${\rm B}(Z\rightarrow ll)$ are 
the leptonic branching fractions of the $W$ and $Z$ bosons.
We extract ${\rm B}(W\rightarrow l\nu)$ from the 
above ratio using a theoretical prediction for $\sigma_W/ \sigma_Z$,
and the precise measurement of ${\rm B}(Z\rightarrow ll)$
from LEP.  We then combine ${\rm B}(W\rightarrow l\nu)$ with 
the leptonic partial width ${\Gamma(W\rightarrow l\nu)}$
to obtain the total width of the $W$ boson, $\Gamma(W)$.

Many of the systematic uncertainties (both experimental and theoretical) that
affect the determination of the individual cross sections 
$\sigma_W 
{\rm B}(W \rightarrow l \nu)$ and 
$\sigma_Z 
{\rm B}(Z \rightarrow ll)$
cancel when calculating the ratio $R$. 
At the present time,
$R$ gives the best determination of $\Gamma(W)$;
direct measurements from fits to the tail of
the transverse mass distribution 
of the $W$ boson are currently four times less 
precise~\cite{CDF1aD}, but require fewer standard model assumptions.

In this paper, we report results
of the  measurement
of the $W$ and $Z$ production cross sections, 
and the extraction of 
$\Gamma(W)$,
using data collected 
in the first
collider run of the D\O\ detector starting in August 1992 and 
ending in June 1993. 
During the run, the Tevatron operated
with a typical instantaneous luminosity of
$4.0 \times 10^{30}$ cm$^{-2}$s$^{-1}$ and a peak luminosity of
$9.7 \times 10^{30}$cm$^{-2}$s$^{-1}$.
D\O\ recorded to tape a total of $\sim 13$ pb$^{-1}$ of data.

\section{The D\O\ Detector}

\begin{figure}[ht]
\centerline{\epsfysize=12cm \epsffile{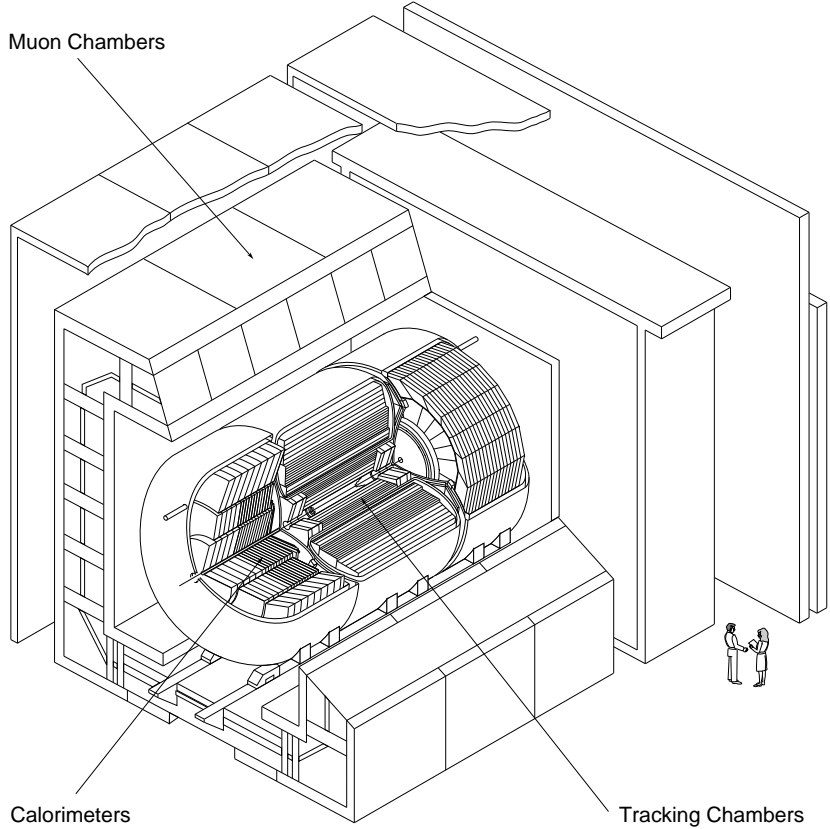}}
\caption{Cutaway isometric view of the D\O\ detector.
\label{figure:figdet1}}
\end{figure}

D\O\ is a multipurpose detector designed to study  $p \bar p$ collisions at
the Fermilab Tevatron Collider. 
It consists of three primary components: a nonmagnetic central tracking
system, a nearly hermetic uranium/liquid-argon calorimeter, and a 
magnetic muon spectrometer.
A cutaway view of the detector is shown in Fig{.}~\ref{figure:figdet1}.
A full description can be found in Ref.~\cite{D0det};
below we give details of the detector relevant to this analysis.

\subsection{Central Tracking System}

The central tracking system consists of four detector subsystems: a vertex
drift chamber (VTX), a transition radiation detector (TRD), a central drift
chamber (CDC), and two forward drift chambers (FDC). The system provides
charged-particle tracking over the pseudorapidity region $|\eta| < 3.2$,
where $\eta = \tanh^{-1} (\cos \theta)$,
$\theta$ is the polar  angle, and $\phi$ is the azimuthal angle. 
Trajectories of charged particles are measured with a resolution of
2.5~mrad  in  $\phi$ and 28 mrad in $\theta$. From these
measurements, the position of the interaction vertex along the beam direction
($z$) can be determined with a typical resolution of 8 mm.
The central tracking system
also measures the ionization of tracks, and can be used
 to distinguish single charged
particles and $e^+e^-$ pairs from photon conversions.

\subsection{Calorimeter}

Surrounding the central tracking system is the calorimeter,
which is divided into three parts: a central calorimeter (CC) and
two end calorimeters (EC).
They each consist of an inner electromagnetic (EM) section, a fine
hadronic (FH) section, and a coarse hadronic (CH) section, housed in a steel
cryostat. The intercryostat detector (ICD) consists of scintillator tiles
inserted in the space between the EC and CC cryostats. The ICD 
improves the energy
resolution for jets that straddle two cryostats. The calorimeter
covers the range $|\eta| < 4.2$.

Each EM section is 21 radiation lengths deep, and 
is divided into four longitudinal
segments (layers). The hadronic sections are 7--9 nuclear interaction lengths
deep, and are divided into four (CC) or five (EC) layers.
The calorimeter is transversely segmented
into pseudoprojective towers
of $\Delta\eta\times\Delta\phi$  = $0.1 \times 0.1$.
The third layer
of the EM calorimeter, in which the maximum energy deposition of 
EM showers is expected,
is segmented twice as finely into cells
with $\Delta\eta\times\Delta\phi$  = $0.05 \times 0.05$. With this fine 
segmentation, the position resolution for electrons above
50 GeV in energy is about 2.5 mm. The energy resolution is
$\sigma(E)/E = 15\%/ \sqrt{E\hbox{(GeV)}} \oplus 0.4\%$
for electrons. For charged pions the resolution is about
$50\%/ \sqrt{E\hbox{(GeV)}}$, and for jets about $80\%/ \sqrt{E\hbox{(GeV)}}$
\cite{D0det}.
From minimum-bias data, for the imbalance in transverse momentum, or 
``missing $E_T$''(see Sec.~\ref{sec:nu}), 
or {\hbox{$\rlap{\kern0.25em/}E_T$}},
the resolution for each component 
({\hbox{$\rlap{\kern0.25em/}E_{x}$}} and {\hbox{$\rlap{\kern0.25em/}E_{y}$}})
is $1.08\hbox{ GeV} + 0.019 (\Sigma E_{T})$, 
where $\Sigma E_{T}$ is the scalar sum of the transverse energies over all
calorimeter cells.

The readout of the individual calorimeters cells are subject to zero
suppression.  They are read out only if the signal is outside of a
two-standard-deviations window centered on the mean of the noise.

\subsection{Muon Spectrometer}

Outside the calorimeter, 
there are muon detection systems covering $|\eta|\leq 3.3$.
Since muons from $W$ and $Z$ boson decays populate predominantly the central
region, this work uses only the wide angle muon spectrometer (WAMUS), which
consists of four planes of proportional drift tubes (PDT) in front of
magnetized iron toroids with a magnetic field of 1.9~T, and two groups of 
three planes each of proportional drift tubes behind the toroids.
The magnetic field lines and the wires in the drift tubes are oriented
transversely to the beam direction.
The WAMUS covers the region $|\eta| < 1.7$ over the entire azimuth, with the
exception of the central region below the calorimeter ($|\eta| < 1$,
$225^{\circ}< \phi  < 315^{\circ}$), where the inner layer is missing to make
room for the support structure of the calorimeter.

The total material in the calorimeter and iron toroids varies between 13 and
19 interaction lengths, making background in the muon chambers from hadronic 
punchthrough negligible.
The D\O\ detector is significantly more compact than previous magnetic 
$p\bar p$
collider detectors~\cite{UA1det,CDFdet}, and the small tracking volume
reduces backgrounds from muons from inflight decays of $\pi$ and $K$
mesons.

The muon momentum $p$ is measured from the muon
 deflection angle in the magnetic
field of the toroid.
The momentum resolution is limited by multiple scattering in the
traversed material, knowledge of the magnetic field, and
measurement of the deflection angle. The resolution in $1/p$
is approximately Gaussian and given by $\sigma(1/p) = 0.18(p-2)/p^2 \oplus
0.008$ (with $p$ in GeV) for the algorithm that was used to select the data
presented here.
The first of the two components in the above resolution function arises 
from multiple-Coulomb scattering in the iron toroids, and is the dominant 
effect for low-momentum muons. 
The second component is from the resolution on 
the measurement of the
muon trajectory.

\section{Particle Identification and Event Selection}

Because it is more  difficult to separate the hadronic decays of
$W$ and $Z$ bosons from the large  background of dijet production,
the cross section analysis uses the  leptonic  decay modes,   
$W\rightarrow   l\nu$ and  $Z\rightarrow ll$ with $l=e$~\cite{yang_grud},
$\mu$~\cite{gerber}.

The leptonic decays 
are characterized by a high-$p_T$ lepton and large 
{\hbox{$\rlap{\kern0.25em/}E_T$}}\ 
or by two high-$p_T$  leptons, for $W$ or $Z$ boson decays, respectively. 
This section describes the identification criteria used for
electrons, muons, and neutrinos in this analysis.

\subsection{Events with Electrons}

Electrons are identified primarily by the presence of an electromagnetic
shower in the calorimeter. A clustering algorithm finds these showers, and
quality criteria are used to pick out electrons and photons,
and thereby reduce backgrounds.
Information from the central tracking system is used to separate electrons
from photons.

This analysis considers electrons in the Central Calorimeter (CC)
defined by $|\eta| \le 1.1$, and the End Calorimeters (EC),
$1.5 \le |\eta| \le 2.5$. The region between the calorimeters
is excluded because of poor resolution.
In the CC, we also exclude electrons within 0.01 radians
in $\phi$ of the crack between adjacent calorimeter modules.

\subsubsection{Clustering Algorithm}

A nearest-neighbor cluster-finding algorithm~\cite{cluster}
 is employed to find the 
electromagnetic energy clusters to be associated with electrons or photons.
In each $\Delta \eta \times \Delta \phi = 0.1 \times 0.1$ tower,
we sum the energies in all layers of 
the EM calorimeters.
We then loop over all such EM energy towers 
with 
$E > 50$~MeV, and search the nearest-neighbor towers for high 
transverse energy. If there are other towers
with $E > 50$~MeV, 
a local connection is made between those neighboring towers. In the next step,
clusters are defined as groups of connected towers.
If the transverse energy of the cluster is greater than $1.5$ GeV,
the cluster is saved for further analysis.
The energy in the EM portion of the calorimeter
is also required to exceed $90\%$ of the total energy of
the cluster, and the energy outside the central tower 
must be less than $60\%$ of the total. Both of these requirements 
are chosen to select clusters corresponding to narrow 
EM particle showers, as expected for electrons or photons.

At this stage, the ``electron'' sample has a very 
large background from QCD processes 
(such as dijet production). 
This is because hadronic showers from QCD jets can sometimes
fluctuate to look like electron or photon showers. 
Several other variables are introduced therefore to clean up 
the electron and photon selections.
These variables 
involve a $\chi^2$ for the shape of the shower, 
the shower's
isolation, and 
the spatial match between the calorimetric shower and
the extrapolated position of some charged track 
emanating from the interaction vertex.

\subsubsection{Covariance Matrix for the Shower}

   Development of electron or photon showers in calorimeters
is characteristically different from that of jets.
The profile of the shower both in the longitudinal
and transverse directions can therefore be used to 
discriminate between signal and QCD background. 
A covariance matrix is constructed to compare the shape of the 
experimentally observed shower with that expected from electrons 
or photons, taking into account the correlations
among energy depositions in all the calorimeter cells in the cluster.  

For a sample of N electrons, a covariance matrix 
is defined as
\begin{equation} M_{ij} = \frac{1}{N} \sum_{n=1}^{N} (x_{ni} - <x_{i}>)
(x_{nj} - <x_{j}>), \end{equation}
where  $x_{ni}$ is the value of the $i$th observable 
for the $n$th electron, and $<x_{i}>$ is 
the sample mean for that observable.  
There are 41 variables in the matrix:
the fractions of shower energy in
EM layers 1, 2 and 4,
the fraction of shower in each cell in 
a $6\times6$ array centered at the hottest tower in EM layer 3,
the logarithm of the shower energy, and
the $z$ position of the event vertex.
A matrix , based on
Monte Carlo simulation of electron showers, is constructed 
for each of the 37 detector towers at different values of $|\eta|$.
The Monte Carlo simulation was tuned to agree with the shower shapes of test
beam electrons.

\begin{figure}[ht]
 \centerline{\epsfysize=12cm\epsffile{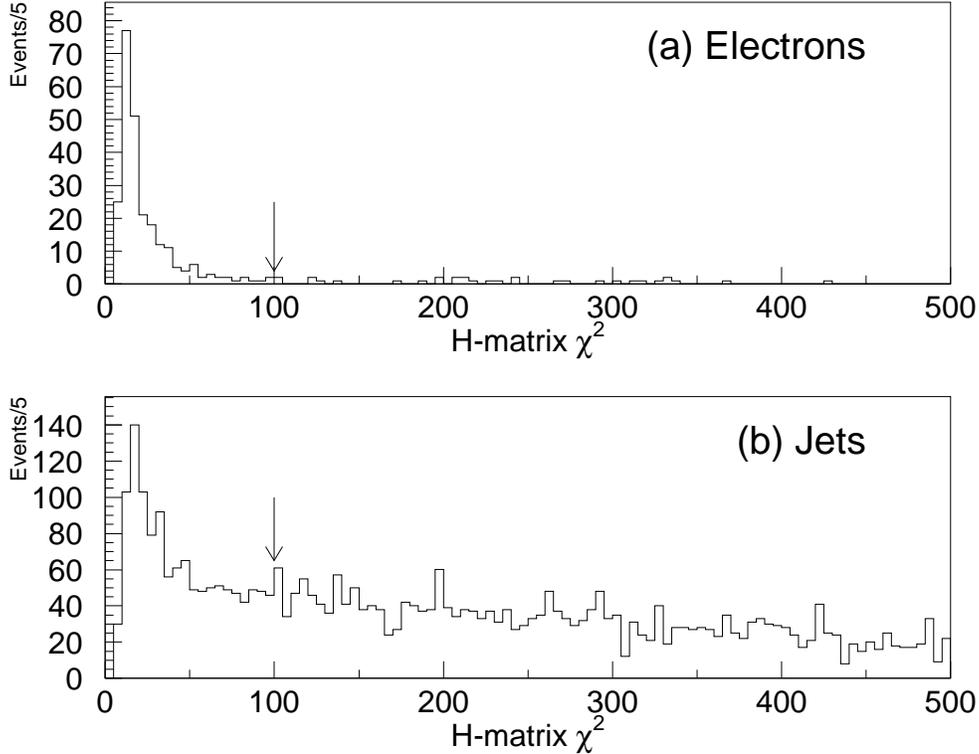}}
 \caption{$\chi^{2}$ distribution for (a) electrons and (b) jets.
The arrows indicate the position of the cutoff used in this analysis
at $\chi^2 = 100$.}
 \label{fig:el_chi2}
\end{figure}

For showers in the data, we calculate the H-matrix $\chi^2$ function:
\begin{equation} 
\chi^{2} = \sum_{ij} (x_{i} - <x_{i}>)H_{ij}(x_{j} - <x_{j}>), 
\end{equation}
where $x_{i}$ is the measured value 
of the $i$th observable, and 
$H = M^{-1}$.
Figure~\ref{fig:el_chi2} shows the distribution of $\chi^{2}$ 
for showers from electron candidates from $Z \to ee$ decays,
and EM clusters in inclusive jet events that are primarily from overlaps
between
charged and neutral particle and $\pi^0$ decays.
The two distributions are clearly different.
Note that  the covariance $\chi^{2}$  parameter will not 
necessarily follow a standard $\chi^{2}$  distribution,
because, in general, the observables defining the matrix
are not normally distributed~\cite{elec_id}.
We require that 
an acceptable electron shower have $\chi^2 \le 100$. 

\subsubsection{Isolation Parameter}

An isolation variable is very useful for discriminating between
background from jets and electrons
from $W$ or $Z$ decay.  Such electrons
usually have very few other particles in their vicinity, while
a jet contains 
many collimated particles 
close to each other.
We therefore reject electron candidates with a significant amount
of energy deposition nearby in the calorimeter.

\begin{figure}[ht]
 \centerline{\epsfysize=12cm\epsffile{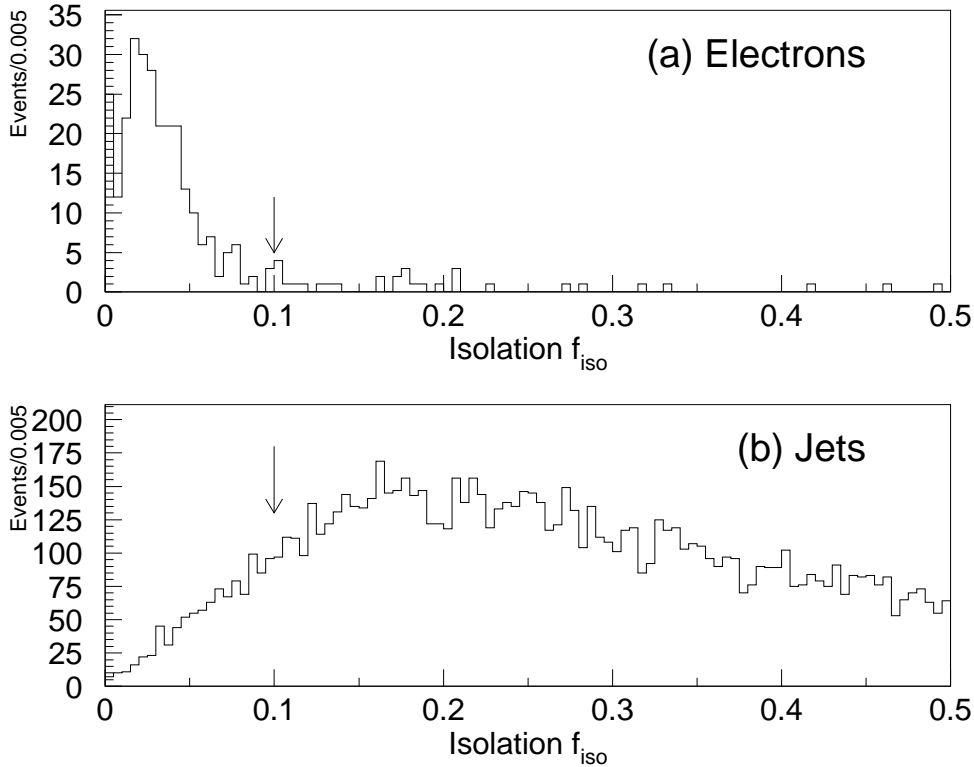}}
 \caption{Distribution in the isolation fraction for 
(a) electrons and (b) jets.
The arrows indicate the cutoff used in this analysis at $f_{\rm{iso}}=0.1$.}
 \label{fig:el_isol}
\end{figure}

The isolation parameter for a cluster is defined by the fraction
of  energy 
in the vicinity of the core towers of that cluster  
\begin{equation} f_{{\rm iso}} = \frac{E_{{\rm tot}}({\cal R}=0.4) - E_{{\rm EM}}({\cal R}=0.2)}
   {E_{{\rm EM}}({\cal R}=0.2)} \end{equation}
where $E_{{\rm tot}}$ is the total energy 
in the calorimeter in a cone with a radius 
${\cal R} = \sqrt{(\Delta \eta)^2 + (\Delta \phi)^2} = 0.4$, and 
$E_{{\rm EM}}$ is the energy in the EM section in a cone with radius $0.2$.
Figure \ref{fig:el_isol} shows the distributions 
found for electrons and jets. 
We require that acceptable 
electrons satisfy $f_{{\rm iso}} < 0.10$.

\subsubsection{Track Matching in the Central Detector}

\begin{figure}[ht]
 \centerline{\epsfysize=12cm\epsffile{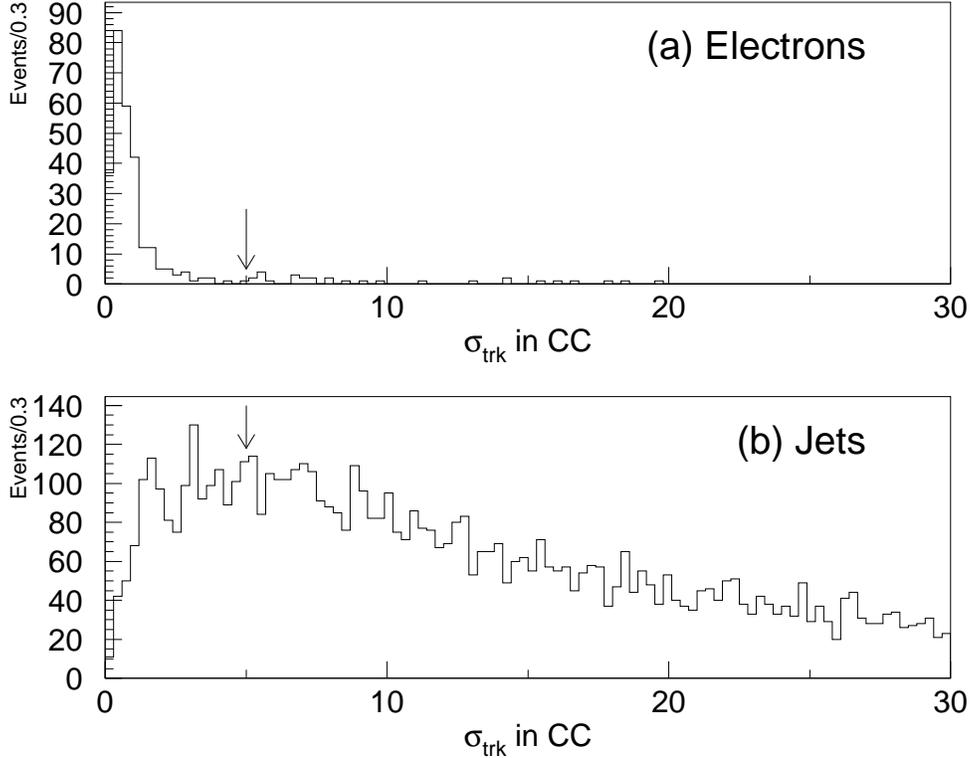}}
 \caption{Significance track matching for (a) electrons in the CC
and (b) jets in the CC.
The arrows indicate the value of the cutoff used in this analysis
at $\sigma_{\rm{trk}}=5$
in the CC.}
 \label{fig:el_match}
\end{figure}

Track information is used to distinguish electrons from 
photons.
A reconstructed track is required to be within 
a $0.1\times0.1$ cone 
pointing towards the centroid of the EM shower. If this requirement is 
satisfied, the cluster is classified as an electron candidate,
otherwise it is considered 
as a photon candidate.
A track significance is defined as a measure of 
the quality of the match between the track and the centroid of 
the shower.   
For the central drift chambers (CDC), it  is defined as: 
\begin{equation} \sigma_{{\rm trk}} = \sqrt{ \left( \frac{\delta z}{\sigma_{z}} \right)^{2} +
 \left( \frac{\delta \phi}{\sigma_{\phi}} \right)^{2} } \end{equation}
while for the forward drift chambers (FDC), it  is defined as:
\begin{equation} \sigma_{{\rm trk}} = \sqrt{ \left( \frac{\delta \rho}{\sigma_{\rho}} \right)^{2} +
 \left( \frac{\delta \phi}{\sigma_{\phi}} \right)^{2} } \end{equation}
where $z,\rho$, and $\phi$ are cylindrical coordinates and 
all the differences  ($\delta$'s) are calculated between the 
extrapolated coordinates
of the track and the centroid of the shower in 
EM layer 3 of the calorimeter.
The standard deviations 
($\sigma$'s) in the denominators are the 
experimental resolutions 
for the corresponding matching parameters.
Figure \ref{fig:el_match} shows the 
distributions in $ \sigma_{{\rm trk}}$ 
for electrons and jets in the CC, and the cutoffs used
to define electrons.

\subsubsection{Energy Scale Calibration}

The absolute EM energy scale of the D\O\ calorimeter
is determined using events in the 
$Z \to e e$ mass peak~\cite{1amwprd}.
An initial calibration was performed based on
test beam studies of electrons and pions in a prototype
calorimeter module. 
These determined that any nonlinearity and energy offsets
of the calorimeter were negligible.
We set the absolute scale by measuring the $Z$
invariant mass peak and scaling our initial result
to the known value of the $Z$ mass~\cite{lepz}.
This correction is determined separately for
each of the three calorimeter cryostats.
The magnitude of the correction ranges from 1\% to 7\%
\cite{escale}.

\subsubsection{Defining Electron Categories}

We define three categories of reconstructed electrons,
referred to as
``tight'', ``standard'',  and ``loose''.
The tight criteria are used to reduce backgrounds as much as 
possible, while the loose criteria are used to obtain a higher
reconstruction efficiency for electrons.  The standard criteria 
are employed in the measurement of electron efficiencies.

Tight electrons are defined as reconstructed EM clusters that:
\begin{itemize}
 \item pass the single-electron trigger (see next section)
 \item have large EM fractions: 
$f_{{\rm EM}} = \frac{E_{{\rm EM}}}{E_{{\rm tot}}} > 0.95$
 \item have H-matrix $\chi^{2} < 100$
 \item are isolated: $f_{{\rm iso}} < 0.10$
 \item have a good matching track, with $\sigma_{{\rm trk}} < 5$ for 
 a CC, and $\sigma_{{\rm trk}} < 10$ for EC.
\end{itemize}
Standard electrons are defined with the same criteria, except for
relaxed requirements on electromagnetic fraction ($f_{{\rm EM}}<0.9$)
and isolation ($f_{{\rm iso}} < 0.15$).
The loose electron definition is the same as that for tight electrons,
with the omission of the 
requirements for the trigger and for a matched track.

\subsubsection{Single-Electron Trigger}

D\O\ uses a multiple-level trigger system.
Common to many triggers 
used in this analysis is the Level-0 trigger,
which requires signals in two hodoscopes
of scintillation counters that are mounted close to the beam region
on the front surfaces of the end calorimeters.
Each analysis uses its own subset of Level-1 (hardware)
and Level-2 (software) triggers.

A single electron trigger is used for both
$W\to e\nu$ and $Z\to ee$ events to benefit from cancelations in
trigger efficiencies when the cross section ratio is determined.
The Level-1 trigger for single electrons demands that 
there be at least one electromagnetic trigger 
tower with transverse energy above 10 GeV
(or 12 GeV for a small fraction of the early data).  
A trigger tower consists of four fixed calorimeter towers,
covering $\Delta\eta \times \Delta\phi = 0.2\times0.2$,
and contains most of the energy of an EM shower.

The Level-2 trigger for electrons searches for the tower 
($\Delta\eta \times \Delta\phi = 0.1\times0.1$) 
that contains the highest energy in the calorimeter,
and then uses the nine ($3\times3$) towers centered on it
to form a cluster. The transverse energy for this cluster 
is required to be greater than 20 GeV in order to pass Level-2.

Level-2 also has minimal
quality cuts on the
shower shape of the cluster. The fraction of the cluster energy in the
EM section must be above a given threshold which is dependent on the
energy of and the position of the cluster in the detector.  The
transverse shape classification is based on the energy deposition
pattern in the third EM layer.  The difference of the energy depositions
in two regions, covering $\Delta\eta\times\Delta\phi = 0.25\times 0.25$
and $0.15\times 0.15$ and centered on the cell with the highest $E_T$,
must be in a window, which depends on the total cluster energy.
Additionally, there is an isolation requirement,
similar to that described above.
The size of the outer cone in the trigger was set to either
0.4 or 0.6, with roughly half the data taken under each condition.

\subsubsection{Criteria on Shower Quality and Electron Kinematics}

Both the $W$ and $Z$ selections require one tight electron as
defined in the previous section. The $W \to e \nu$ selection
requires a tight electron with $E_T \ge 25$ GeV, ${\hbox{$\rlap{\kern0.25em/}E_T$}} \ge 25$ GeV,
and no second high $E_T$ electron electron.
A total of 10338 candidate events satisfy these requirements.  

For $Z$ events, there is an additional requirement of a 
second loose electron with $E_{T} > 25$~GeV. 
Also, the invariant mass of the two electrons ($M_{ee}$)
is restricted to the range 75--105 GeV.
A total of 775 $Z$ candidates satisfy all the criteria.

 Distributions of the transverse mass for the $W\to e\nu$ events and the 
invariant mass 
for the $Z\to ee$ events are shown in Fig.~\ref{fig:mass_plots}.
The transverse mass, $M_T$, is defined by 
$M_T^2=2E_T^l\rlap{\kern0.25em/}E_T(1-\cos\Delta\phi_{l\nu})$, where 
$\Delta\phi_{l\nu}$ is the azimuthal separation between the charged lepton
and the missing transverse energy vector.

\subsection{Events with Muons}

Muons are identified by reconstructing a track from hits in the muon
PDTs. The track is confirmed using information
from the calorimeter and the central detector.
Part of the confirmation is
a ``global fit'', which uses not only hit positions from the PDTs
but also those from the other detector systems.
It consists of a
fit to the position of the primary vertex, a track in the 
central detector, and the muon track before and after
the toroid.  The seven parameters in the fit include
four for the position and angle 
of the track before the calorimeter (in both the bend and non-bend views), 
two describing the effects of the multiple scattering
in the calorimeter; and one for the reciprocal of the momentum ($1/p$).
The momentum is therefore determined by the deflection in the magnetized iron
with a correction for the expected energy loss in the calorimeter
\cite{geant}.

This analysis uses muons contained entirely in the central
WAMUS detector ($|\eta| \le 1.0$).
To obtain a reliable momentum measurement,
the minimal value of the integral
of the magnetic field along the muon track is required to be
$\int B 
dl \ge 2$ Tm. 
Although this reduces significantly the acceptance for muons,
it also eliminates a potential background from punchthrough.
In the regions of low $\int B 
dl$, the D\O\ detector has only  about 9
interaction lengths, while it has typically 13--18
interaction lengths elsewhere.
This requirement therefore provides a good momentum measurement, and a
cleaner sample of muons because of the greatly reduced probability
of hadron
punchthrough for tracks from the calorimeter.

\subsubsection{Confirmation from Calorimeter and Central Detector}

\begin{figure}[ht]
 \centerline{\epsfysize=12cm\epsffile{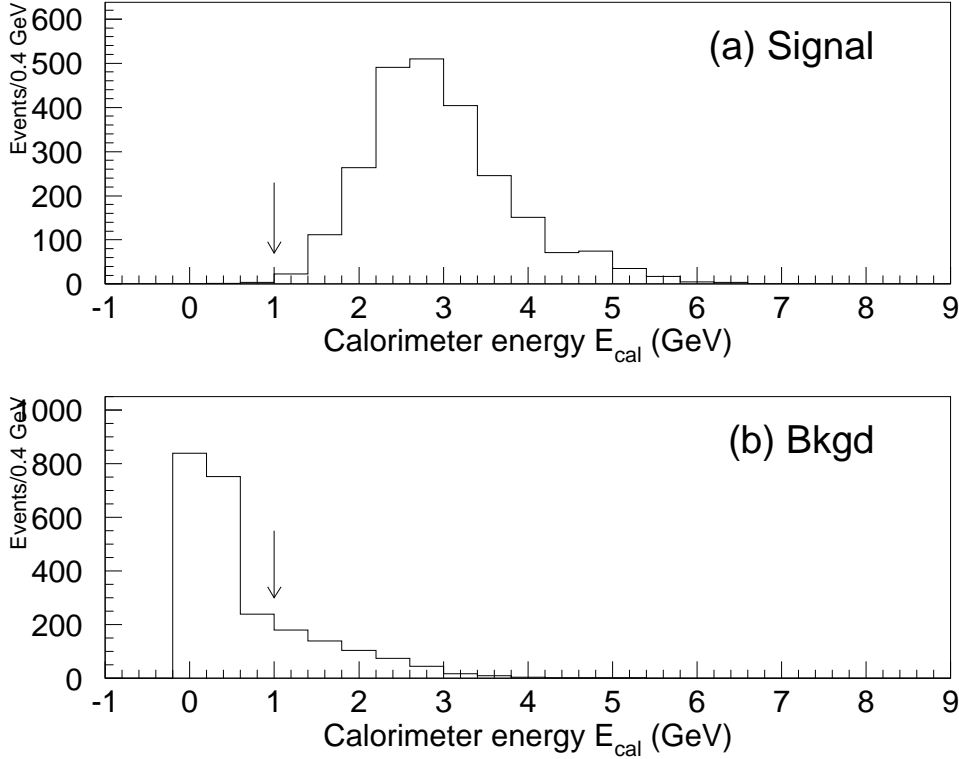}}
 \caption{Calorimeter energy for (a) good muons and (b) background.
The arrows indicate the cutoff used in this analysis at $E_{\rm{cal}}=1$ GeV.}
 \label{fig:mu_mip}
\end{figure}

Candidate muon tracks 
found in the PDTs must be confirmed by the presence of 
energy deposited along their trajectories in the calorimeter.
This reduces background from
cosmic ray muons and from random combinations of
PDT hits.
We require a sum of at least 1.0 GeV of energy deposition 
in the cells of the extrapolated trajectory of the muon  
and in the two nearest-neighbor cells. A muon typically deposits
$\sim 3$ GeV in this volume. 
Figure~\ref{fig:mu_mip} shows the energy deposited in the calorimeter 
for good muons and for background.

Another effective way to reduce background is to require a track match 
between that of the muon system track and the central detector.
We require that there be a CD track associated with the muon,
and that the angles between the two tracks match to within
$ \Delta\phi$ (muon track, CD track) $ \leq 0.04$ radians,
and
$ \Delta\theta$ (muon track, CD track) $ \leq 0.12$ radians.

\subsubsection{Rejection of Cosmic Rays}

Additional rejection of
cosmic-ray muons and
background from combinations of random PDT hits 
is provided by 
requiring  small 
impact parameters of the muon track relative to the interaction vertex, and 
correct drift time relative to beam crossing.

The impact parameters for muon tracks, both in the bend and 
nonbend views, are calculated by extrapolating the muon trajectory
inside the toroids 
back to the primary interaction vertex. 
To be acceptable, the muon track has to point to the primary
interaction vertex within 15 cm in the bend view and 20 cm in the
nonbend view.

The muon timing is determined
by allowing the drift times of all the muon hits
to vary coherently.  The time interval
$t_0^f$ is defined as the offset between the beam crossing time 
and the time that gives the best fit for the muon track.
Because they are produced in coincidence
with beam crossings,  
prompt muons have a $t_0^f$
distribution that peaks at zero 
However, cosmic rays arrive at random times.
To have most of its PDT hits recorded,
a muon has to arrive within
about $\pm400$ ns of the beam crossing time (the total PDT drift time 
is $\approx 750$ ns).
Due to the finite rise time of the trigger signals,
the probability for accepting cosmic rays 
is enhanced somewhat for early arrivals 
($t_0^f > 0$). 
To reject cosmic rays we require  $t_0^f \leq 100$ ns.

\begin{figure}[ht]
 \centerline{\epsfysize=12cm\epsffile{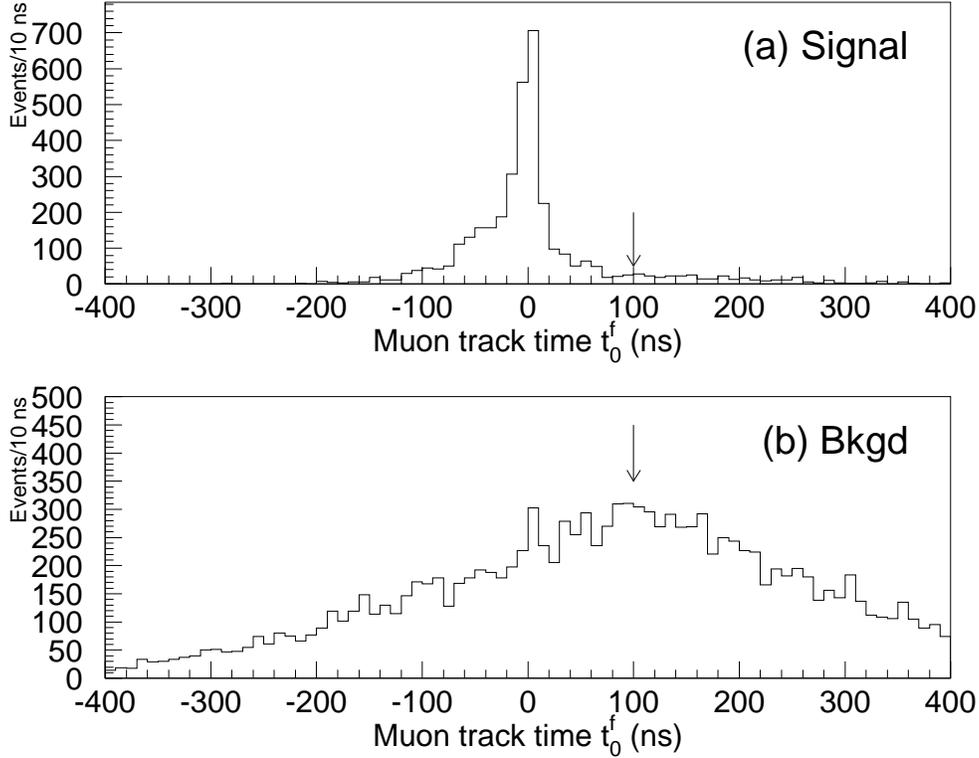}}
 \caption{The $t_0^f$ distributions for (a) good muons and (b) background.
The arrows indicate the cut used in this analysis at $t_0^f = 100$ ns.}
 \label{fig:tfloat}
\end{figure}

Figure~\ref{fig:tfloat} shows the distribution of $t_0^f$ for signal and
background samples.
The background was obtained by selecting collider events containing 
two isolated high $p_T$ muons that are back-to-back 
($\Delta\theta(\mu_1,\mu_2)>170^{\circ}$ and 
$\Delta\phi(\mu_1,\mu_2)>160^{\circ})$.
Such a sample is dominated by cosmic-ray muons.

\subsubsection{Global Fit to Muons}

\begin{figure}[ht]
 \centerline{\epsfysize=12cm\epsffile{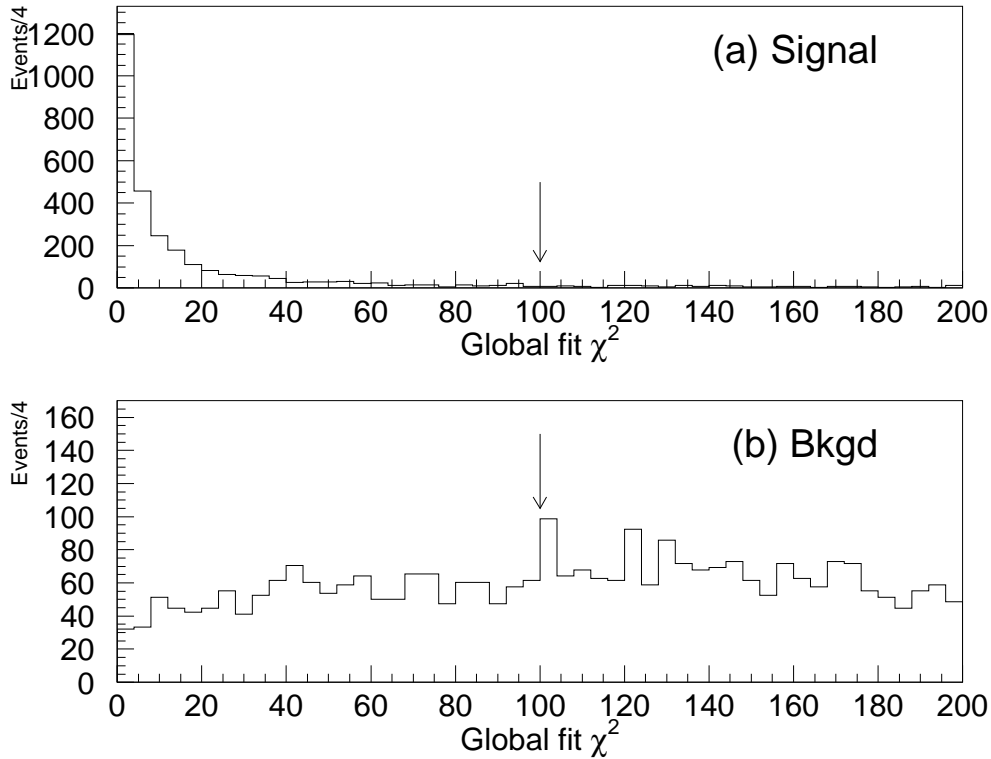}}
 \caption{Global-fit $\chi^2$ distribution for (a) good muons and (b)
background.
The arrows indicate the cutoff used in this analysis at $\chi^2 = 100$.}
 \label{fig:gfit}
\end{figure}

The quality of the global fit
of a muon track 
is characterized by the value of the $\chi^2$ for the fit, 
and depends on the parameters of the muon system as well as on
those of the tracking system.
By using the additional information, we are able to reduce
the backgrounds from cosmic rays and 
from random combinations of PDT hits.
The $\chi^{2}$ distributions for a signal 
and background are plotted in
Fig.~\ref{fig:gfit}.  To accept an event,
we require a fit with $\chi^2 \le 100$.

\subsubsection{Muon-Isolation Parameters}

A background that is not affected by the above criteria
is that from  QCD jet production.
These events can have muons resulting from semileptonic decay of produced
hadrons (e.g., $b\bar{b}$ events).  Such muons are usually associated with
jets, while muons from $W$ or $Z$ decay are most often isolated.
We reduce the QCD background by imposing specific requirements on
the calorimeter energy deposited within ${\cal R} = 0.2$ and 0.6 of the muon.

We define the variable ${\cal I}_2$
as the difference 
between the calorimeter energy observed in cells traversed by the muon 
(including the two nearest-neighbor cells within ${\cal R}=0.2$ of
the muon)
and the expected contribution from the muon ionization,
divided by the uncertainty in the expected energy loss in the calorimeter:
\begin{equation} {\cal I}_2 = 
{{E_{\rm tot}({\cal R}=0.2) - E({\rm expected})} \over
{\sigma_{E({\rm expected})}}}. \end{equation}
The expected energy loss
is determined from the GEANT~\cite{geant}
simulation of the D\O\ detector.
We also define the variable ${\cal I}_6$ as
%
\begin{equation} {\cal I}_6 = E_{{\rm tot}}({\cal R}=0.6) - E_{{\rm tot}}({\cal R}=0.2). \end{equation}

\begin{figure}
 \centerline{\epsfysize=15cm\epsffile{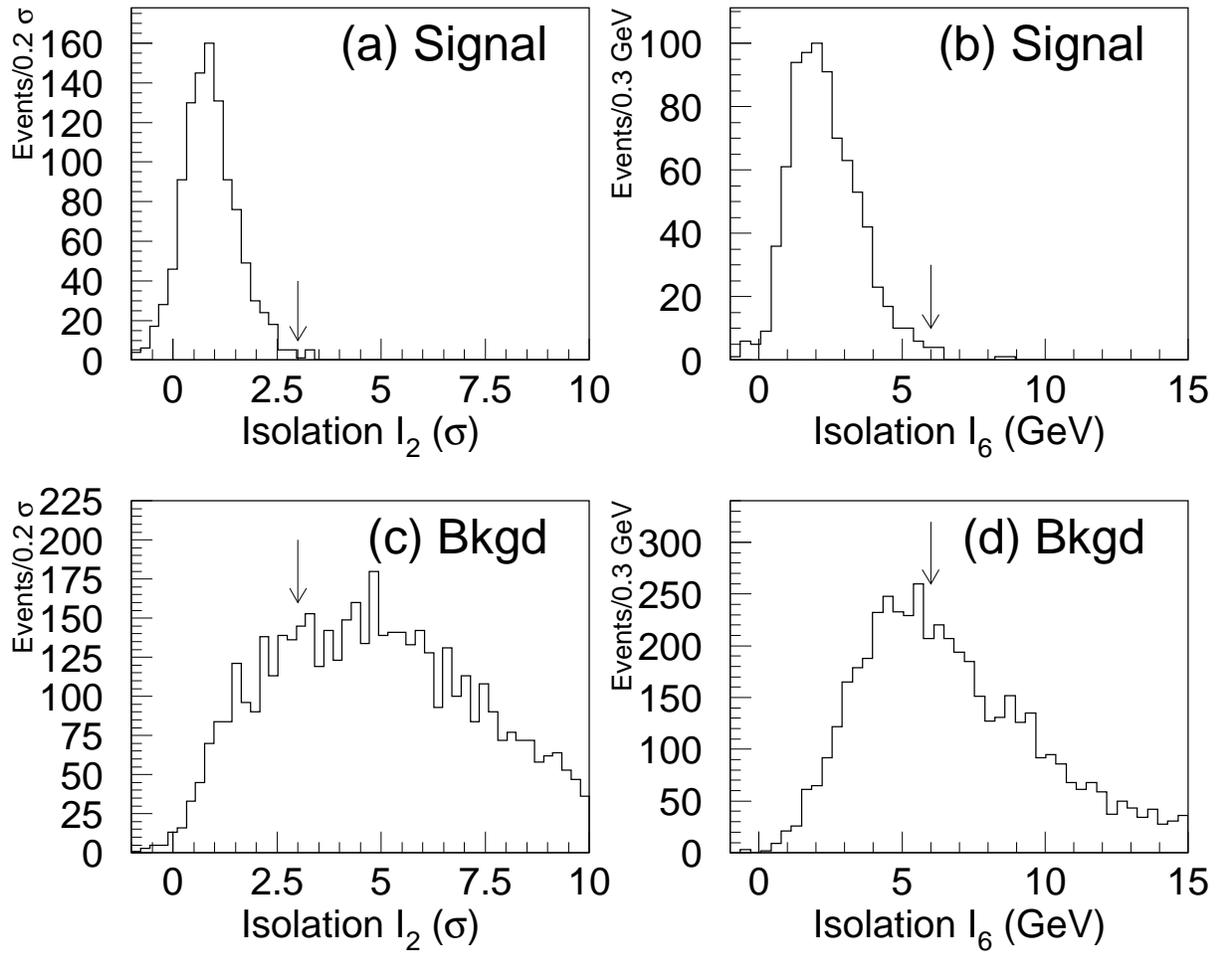}}
 \caption{Isolation distributions for (a) and (b) isolated muons,
 and (c) and (d) non-isolated muons.
The arrows indicate the cutoffs used in this analysis:
$I_2 = 3$ and $I_6 = 6$ GeV.}
 \label{fig:mu_isol}
\end{figure}

Figure~\ref{fig:mu_isol} shows the distributions of ${\cal I}_2$ and 
${\cal I}_6$ for samples of isolated and nonisolated muons.
The isolated muons are from a subset of $W \to \mu \nu$ candidates
with no jets opposite the muon in $\phi$ and the nonisolated muons 
are from events with muons in the range $10 < p_T < 15$ GeV,
a sample dominated by heavy quark decay.
We reduce the QCD background significantly by
requiring that ${\cal I}_2$ $\leq 3$
and ${\cal I}_6$ $\leq 6$ GeV.

\subsubsection{Definitions of Muon Quality}
\label{sec:muid}

We define a ``tight'' muon as a reconstructed track in the PDTs
that has:
\begin{enumerate}
\item calorimeter confirmation with energy in central and nearest-neighbor
cells of $> 1$ GeV
\item a track match in the central detector
\item a successful global fit, with $\chi^2 \le 100$
\item isolation requirements ${\cal I}_2<3$ 
and ${\cal I}_6<6$ GeV
\item no back-to-back muon tracks (or PDT hits). 
\end{enumerate}
A ``loose'' muon is not required to satisfy criteria 2-5.

\subsubsection{Single-Muon Trigger}

The single muon trigger requires a high-$p_T$ WAMUS muon candidate
at both Level 1 and Level 2. The muon Level 1 system has two sublevels
of hardware. The first sublevel passes events if there are PDT hits
within a wide road ($\approx 60$ cm), 
equivalent to a $p_T$ cutoff of 5 GeV.
The second sublevel searches in narrower
roads ($\approx 30$ cm), equivalent to a $p_T$ cutoff of 7 GeV.
The Level 2 software trigger has pattern recognition,
and accepts muons passing $p_T^{\mu}\geq15$ GeV.
Loose quality 
criteria are also applied at Level 2.

Cosmic ray muons are suppressed at Level 2 if there is evidence of  a single
muon penetrating the entire detector. Muon candidates with a track in the
opposite muon chambers within 20$^{\circ}$ in $\phi$ and 10$^{\circ}$
in $\theta$ are rejected, as are those candidates with PDT chamber hits
on the opposite side within 60 cm (roughly 5$^{\circ}$) of the projected
muon track.

\subsubsection{Muon Kinematic and Quality Criteria}

The $W \to \mu \nu$ offline selection requires one tight muon
with $p_T \ge 20$ GeV\ and ${\hbox{$\rlap{\kern0.25em/}E_T$}} \ge 20$~GeV. 
The $Z \to \mu \mu$ offline selection requires
at least one tight and one loose muon. Both muons
must have $p_T \ge 15$ GeV\ and at least one has 
to have $p_T \ge 20$~GeV.
To reject the cosmic ray background, 
we require either $\Delta \phi \leq 160^\circ$
or $\Delta\theta \leq 170^\circ$ between the two muons.
To eliminate low-mass dimuon pairs it is required that 
$\Delta \phi \geq 30^\circ$.
Any event satisfying the $Z$ criteria is removed from the $W$ sample.

\begin{figure}
\centerline{\epsfxsize=17.0cm \epsffile{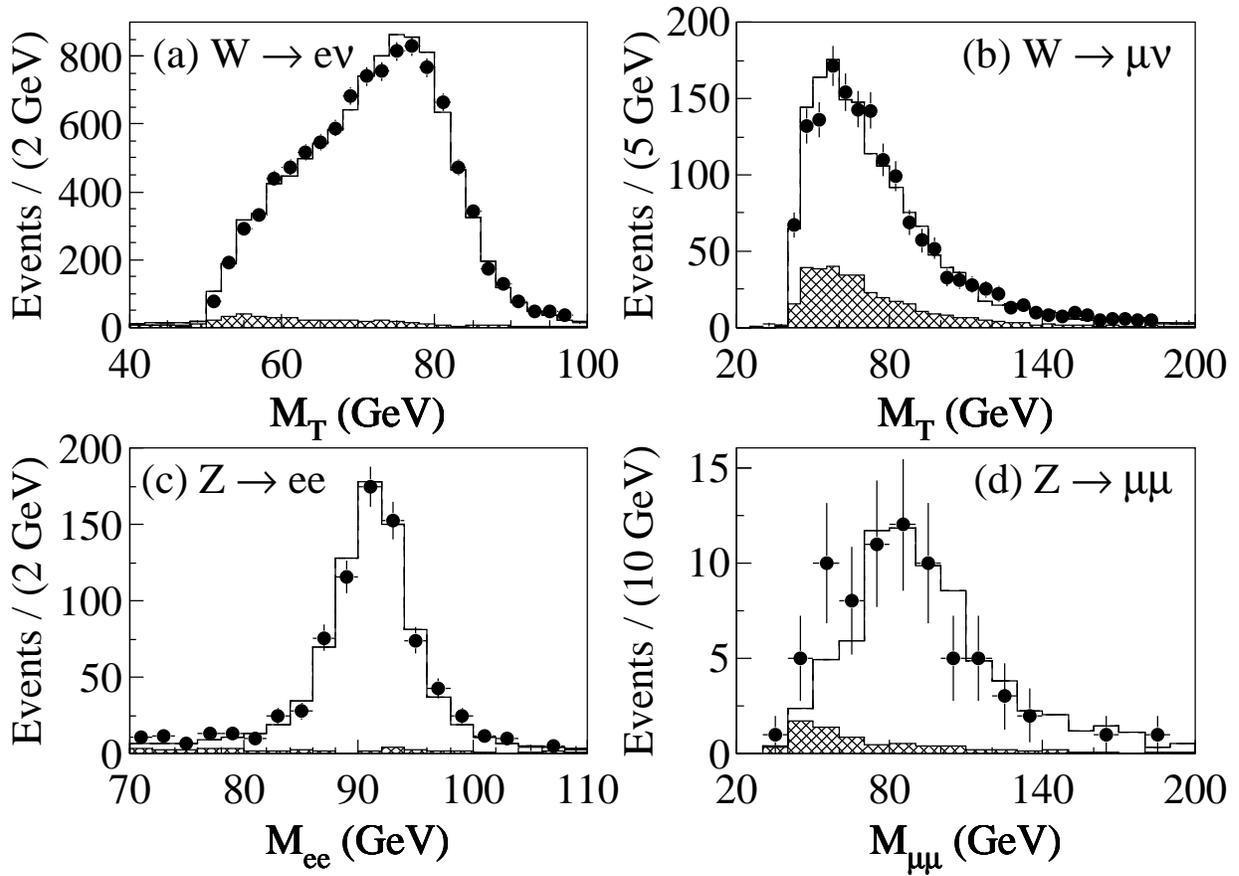}}
 \caption{Transverse mass distributions 
in the final (a) $W\to e\nu$ and (b) $W\rightarrow\mu\nu$ samples
and invariant mass distributions 
in the final (c) $Z\to\ ee$ and (d) $Z\rightarrow\mu\mu$ samples.
The points are the data, the hatched regions correspond to the estimated
backgrounds, and the histograms are the sum of signal (Monte
Carlo) and background.}
 \label{fig:mass_plots}
\end{figure}

   Distributions of the transverse mass for
$W$ events and the dimuon invariant mass 
for $Z$ events are shown in Fig.~\ref{fig:mass_plots}.

\subsection{Neutrino Identification}
\label{sec:nu}

Neutrinos are identified in the D\O\ detector by the presence of
missing transverse energy ({\hbox{$\rlap{\kern0.25em/}E_T$}}).
We define
\begin{equation} {\hbox{$\rlap{\kern0.25em/}E_T$}} = 
\sqrt{({\hbox{$\rlap{\kern0.25em/}E_T$}}_x)^{2}+({\hbox{$\rlap{\kern0.25em/}E_T$}}_y)^{2}}\end{equation} 
where
\begin{equation} {\hbox{$\rlap{\kern0.25em/}E_T$}}_x = - \sum_{i} E_{i}\sin\theta_{i}\cos\phi_{i}, \end{equation}
\begin{equation} {\hbox{$\rlap{\kern0.25em/}E_T$}}_y = - \sum_{i} E_{i}\sin\theta_{i}\sin\phi_{i}\end{equation}
where {\it i} runs over all calorimeter cells with readout signals
after zero suppression,
and $E_{i}$ is the energy deposited in the 
$i$th cell, with $\theta_{i}$ and $\phi_{i}$ as the polar and 
azimuthal angles of that cell, respectively.
If there are muons in the event, we subtract the $p_T$
of the muons as follows:
\begin{equation} {\hbox{$\rlap{\kern0.25em/}E_T$}}_x = - \sum_{i} E_{i}\sin\theta_{i}\cos\phi_{i} - p_{Tx}, \end{equation}
\begin{equation} {\hbox{$\rlap{\kern0.25em/}E_T$}}_y = - \sum_{i} E_{i}\sin\theta_{i}\sin\phi_{i} - p_{Ty}. \end{equation}

The resolution of the missing transverse energy is affected by many factors,
such as statistical
energy fluctuation in the calorimeters, energy lost in
and around the beam pipe and cracks in the central calorimeter,
signal fluctuations caused by the uranium radioactivity,
random and coherent electronic noise. 

\begin{figure}[ht]
 \centerline{\epsfysize=10.0cm \epsffile{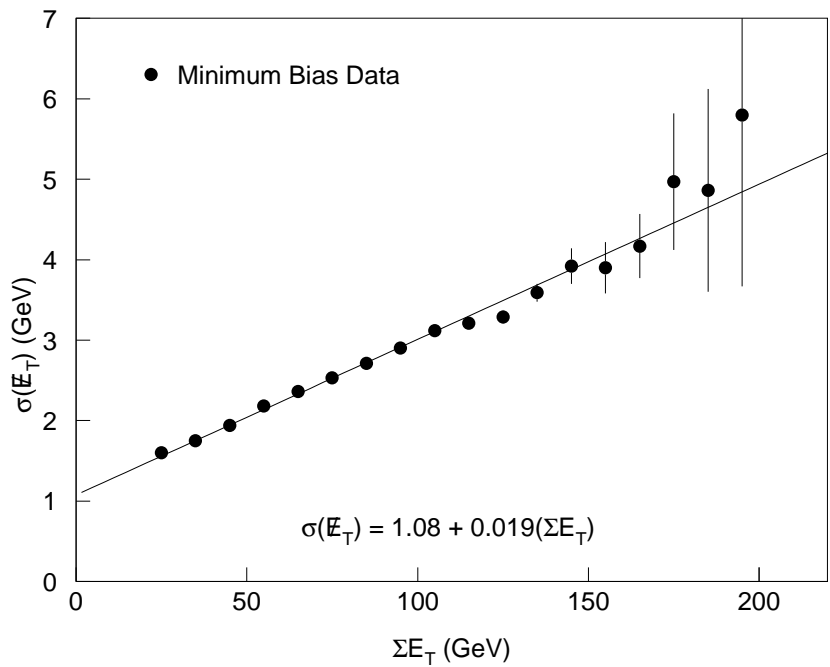}}
 \caption{Resolution in missing transverse energy for minimum bias data.}
 \label{fig:met_res}
\end{figure}

Since we have a nearly hermetic calorimeter with good energy resolution,
we also obtain very good {\hbox{$\rlap{\kern0.25em/}E_T$}}\ resolution.
A global quantity called the scalar transverse energy,
defined as
\begin{equation} \sum E_{T} = \sum_{i} E_{i}\sin\theta_{i} \end{equation}
is used to parameterize the resolution as
\begin{equation} \sigma_{{\scriptscriptstyle{\hbox{$\rlap{\kern0.25em/}E_T$}}}} 
= a + b
   \sum E_{T} \end{equation}
with $a = 1.08$ GeV and $b=0.019$ obtained from minimum-bias data.
Figure~\ref{fig:met_res} shows the dependence of the
{\hbox{$\rlap{\kern0.25em/}E_T$}}\ resolution on $\sum E_{T}$.

\section{Backgrounds}

Backgrounds to $W$ and $Z$ events can be into two groups:
those from ``fake'' leptons, whose levels are estimated from data,
and those from ``physics'' processes that contain
true isolated high-$p_T$ leptons and true {\hbox{$\rlap{\kern0.25em/}E_T$}}. 
The contributions
for the latter sources are estimated from Monte Carlo samples.

Electron background stems primarily from jets and direct photons passing
our electron criteria.
Muon background consists mainly of cosmic-ray muons, random hits in the muon
chambers that form a track, and muons from heavy quark
decays. The inherent background processes,
common to both lepton channels, are
$W \rightarrow \tau \nu \rightarrow l \nu \nu \nu$,  
$Z \rightarrow l l$, 
$Z \rightarrow \tau \tau \rightarrow l l \nu \nu$, 
and Drell-Yan production of $\l^+\l^-$ pairs. 

\subsection{Backgrounds to $W \rightarrow l \nu$}

\subsubsection{QCD background to $W \to e \nu$}

A multijet event  
can be misinterpreted as a $W \rightarrow e \nu$\ in our detector, 
for instance,
if one of the jets fluctuates to have a high electromagnetic content
and passes the electron selection requirements,
while another jet loses energy in the cracks of the detector, or its energy
is otherwise mismeasured, to yield missing transverse energy, 
thereby faking a neutrino.
  
\begin{figure}[ht]
\centerline{\epsfysize=10cm\epsffile{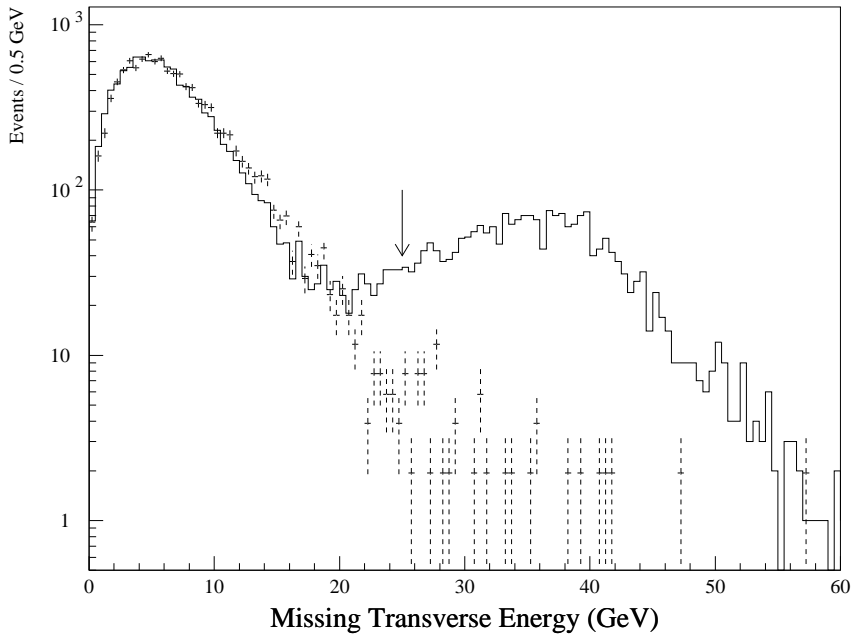}}
 \caption{The {\hbox{$\rlap{\kern0.25em/}E_T$}} distribution for 
the sample of $W \to e \nu$ events (solid histogram). 
The dotted points with error bars represent the 
multijet QCD background 
normalized in the low {\hbox{$\rlap{\kern0.25em/}E_T$}}\ region (0--10 GeV)
to the signal sample.
The arrow indicates the cutoff for the final selection of 
{\hbox{$\rlap{\kern0.25em/}E_T$}} $> 25$ GeV.}
 \label{fig:el_met_bkgd}
\end{figure}

We study this background through the 
{\hbox{$\rlap{\kern0.25em/}E_T$}}\ distribution
of tight-quality electrons prior to the imposition of any
{\hbox{$\rlap{\kern0.25em/}E_T$}}\ criteria.
Figure~\ref{fig:el_met_bkgd} shows that there are two peaks
in the data. 
The peak in the low-{\hbox{$\rlap{\kern0.25em/}E_T$}}\ region is mostly 
due to jet events, and
the second peak is dominated by true $W \rightarrow e\nu$ decays.

We also consider a sample of QCD dijet events,
those for which the electron candidate fails the isolation
criterion (i.e., we require that there be some energy deposition around 
the ``electron", which is presumably due to the rest of the 
remnants of the jet).
Since isolation and {\hbox{$\rlap{\kern0.25em/}E_T$}}\ criteria
are not correlated, the {\hbox{$\rlap{\kern0.25em/}E_T$}}\ spectrum for the dijet events in
this sample and in the tight-electron sample should be the same. 
We therefore normalize the two samples in the low-{\hbox{$\rlap{\kern0.25em/}E_T$}}\ region, 
and extrapolate to find the number of background events
under the $W$ peak passing the {\hbox{$\rlap{\kern0.25em/}E_T$}}\ cutoff 
of 25 GeV.
Figure~\ref{fig:el_met_bkgd} shows the second background
sample normalized to the tight-electron sample. Since the
background falls rapidly with {\hbox{$\rlap{\kern0.25em/}E_T$}} , 
there are very few events
that pass the cutoff.

We consider the QCD background to $W$ decays separately 
for electrons found in the CC and EC calorimeter.
The events are further subdivided into two groups, to
take into account two variants used in 
the electron trigger
(with isolation radius 0.4 and 0.6).
For each of these data subsets, the background
sample is normalized to the signal sample in the region 0--10 GeV in 
{\hbox{$\rlap{\kern0.25em/}E_T$}},
to determine the number of events in the background samples with 
${\hbox{$\rlap{\kern0.25em/}E_T$}} >$ 25 GeV. 
We find the background to electrons in the CC to be $\sim 3\%$ 
and in the EC to be
$\sim 4\%$. The overall jet background in the $W\to e\nu$ data sample is 
$(3.3 \pm 1.7) \%.$
The uncertainty has changed with respect to the original Letter 
\cite{D0PRL} as a result
of additional studies of the {\hbox{$\rlap{\kern0.25em/}E_T$}}
distribution \cite{d0ptW}

\subsubsection{QCD background to $W \to \mu \nu$}

\begin{figure}[ht]
\centerline{\epsfysize=10cm\epsffile{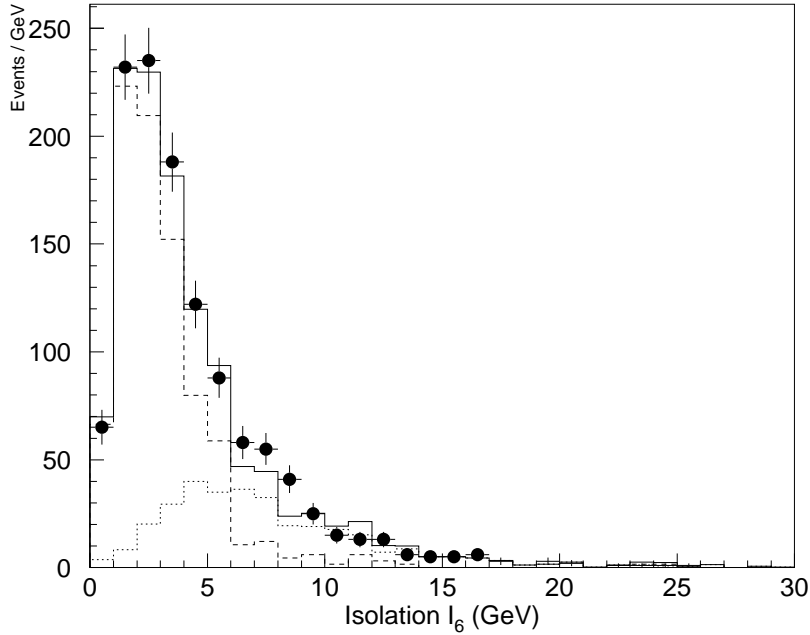}}
\caption{The distribution in the isolation $\cal{I}$$_6$
for the $W \to \mu \nu$ sample (data points), 
fitted to a linear sum of the signal and background. 
The dashed histogram corresponds to isolated muons, the dotted 
histogram shows muons
originating from QCD processes, and the solid histogram represents their sum.}
\label{fig:mu_halo_bkgd}
\end{figure}

The QCD background to $W \to \mu \nu$ events consists of 
muons from decays of particles associated with jets.
Most such muons fail
our isolation criteria.
We estimate the background by fitting the 
observed distribution in 
energy
deposition ${\cal I}_6$ 
(without imposing any isolation criteria)
to a sum of distributions expected for 
isolated and for nonisolated muons
(see Section III.B.4).
The fit to a
linear sum of signal and background 
to the data 
is shown in Fig.~\ref{fig:mu_halo_bkgd}.
After applying the two isolation criteria,
the QCD contamination in the final $W \rightarrow \mu \nu$ sample is
$(5.1 \pm 0.8) \% .$

\subsubsection{Backgrounds to $W \to \mu \nu$ from Cosmic-Ray Muons and 
Random PDT Hits}

The background from cosmic rays and random PDT hits is estimated from 
the $t_0^f$ distributions. 
Since neither background is beam-associated, 
there should be no correlation between the best time for
the fit of the track and the beam crossing.
The prompt distribution is obtained from a 
sample of muons with $p_T >5$ GeV, and very tight quality criteria: 
a matching track in the central detector, tight global fit $\chi^2$,
and sufficient energy deposition 
around the muon trajectory to ensure that it is part of a jet.
The background fraction is determined by fitting the data sample to a
linear sum of signal and background $t_0^f$ distributions. 
The contamination from cosmic rays and random hits
in the final $W \rightarrow \mu \nu$ sample is
estimated to be 
$(3.8 \pm 1.6) \%.$

\subsubsection{Punchthrough and $\pi/K$ Backgrounds to $W \to \mu \nu$}

For  $ p_T >12$ GeV, the background originating from $\pi/K$ decays
is estimated to be an order of magnitude smaller than
that from $b$ decays~\cite{pi_K}. 
The rate from punchthrough is expected to be yet another order of magnitude
lower.
The reason for the low rate is the great 
thickness of the
calorimeter and iron toroid systems at D\O, combined with the fact that the
momentum measurement is made after most of the material has been traversed.
The background contamination from these sources in the 
$W \rightarrow \mu \nu$ and $ Z \rightarrow \mu \mu$  samples
is therefore 
negligible.

\subsubsection{$W \rightarrow \tau\nu \rightarrow l\nu\nu\nu$ Backgrounds to $W \to l \nu$}

The process $W \rightarrow \tau\nu \rightarrow l\nu\nu\nu$ is
experimentally indistinguishable from the signal.  Therefore the only
means for reducing this background is through differences in
kinematics.  Since the background (charged) lepton comes from the
decay of a $\tau$, it will have a much softer $p_T$ distribution than
from direct $W$ decay.
Just the standard kinematic requirements keep this background to a
moderate level.

We use Monte Carlo simulations to calculate the geometric and
kinematic acceptance $A_{\tau}$ of $W \rightarrow \tau\nu \rightarrow
e(\mu)\nu\nu\nu$.  Accounting for the $\tau \to e(\mu)$ branching fraction,
we find the overall $\tau$ background in the electron channel to be
$(1.8 \pm 0.2) \%$, and in the muon channel, it is $(5.9 \pm 0.5) \%.$
The background in the electron channel is lower because the energy
resolution for electrons is better than for muons and the $p_T$
cutoffs are higher in the electron analysis.

\subsubsection{$Z \rightarrow ll$ Backgrounds to $W \to l \nu$}

One of the two leptons from $Z$ decay can escape detection or be
poorly reconstructed in the detector and thereby simulate the presence
of a neutrino, and contribute to the $W \to l \nu$ data sample.
Assuming $\sigma(p\bar{p} \rightarrow Z \rightarrow ee)/
\sigma(p\bar{p} \rightarrow W \rightarrow e\nu)$ is $0.10 \pm 0.01$,
we find from a Monte Carlo simulation that this background fraction is 
$(0.6 \pm 0.1) \%$ for electrons, 
and $(6.5 \pm 0.5) \%$ for muons.  The electron background is lower
because of the greater hermeticity of the calorimeter for electrons
compared to WAMUS for muons.

\subsubsection{$Z \rightarrow \tau\tau \to l \nu \nu l \nu \nu$ Backgrounds to 
   $W \rightarrow l \nu$ }

The process $Z \rightarrow \tau\tau$ has the same rate as $Z
\rightarrow ll$, which is already ten times smaller than the rate of
$W$ production. Each electron from $\tau$ decay has the soft $p_T$
spectrum mentioned for the case of $W \rightarrow \tau\nu$. This
background is therefore doubly suppressed. For the muon channel we
estimate the background to be $(0.8 \pm 0.2) \%,$ and for the electron
channel it is negligible.

\subsection{Backgrounds in the $Z \to l l$ Sample}

\subsubsection{QCD background to $Z \to e e$}

The background to $Z \rightarrow ee$ consists mainly of QCD jet
production, where the jets are misidentified as electrons.  Because
the invariant mass distribution of two electrons from the $Z$ decay
has a well defined resonance peak, and the background has only a weak
dependence on the $ee$ mass, we use the shape of the mass distributions to
estimate background.  We fit a theoretical $Z/\gamma$ line shape and
the experimentally determined shape of the QCD background (see below)
to the data, and determine the absolute normalization of the QCD
background through this fit.

The invariant mass distribution of the QCD background is obtained
from data. We can approximate the 
``two-electron'' mass spectrum from jets 
with $p_T > 25$ GeV,  for the mass range 65--250 GeV,
by an exponential function:
\begin{equation} f_{{\rm dijet}} (m) \propto e^{- 0.0237m}. \end{equation}

A second QCD contribution arises from direct-photon events with
associated jets. Here the jet fragmentation fluctuates sufficiently
for the jet to be reconstructed as an electron, while the photon is
mistaken as a loose electron (only failing the track match).  Again we
can describe this ``dilepton'' mass spectrum by the following
exponential function for the same mass range as above:
\begin{equation} f_{\gamma {\rm jet}} (m) \propto  e^{- 0.0345m}.
\end{equation}

  We use the PYTHIA Monte Carlo~\cite{pythia} to generate the complete
$Z/\gamma$ line shape, including QED radiation from electrons, and we
also simulate the energy resolution of the detector.

Using a maximum likelihood fit to the dielectron invariant mass
spectrum, we determine the fraction of events in the data sample that
can be attributed to QCD background.  We find a total background of
$(2.8 \pm 1.4) \%$, where the error includes statistical as well as
systematic uncertainty to account for the sensitivity to the mass
window used in the fit (71 or 75 GeV to 111 or 121
GeV).

\subsubsection{QCD Background to $Z \to \mu \mu$}

For the $Z \rightarrow \mu \mu$ sample, the background is estimated in
a similar fashion to that for the $W \to \mu \nu$ background, by fitting
the calorimeter energy distribution ${\cal I}_6$.  The QCD background
in the final $Z \rightarrow \mu \mu$ sample is estimated to be $(2.6
\pm 0.8) \%.$

\subsubsection{Backgrounds to $Z \to \mu \mu$ from Cosmic-Ray 
Muons and Random PDT Hits}

The backgrounds from cosmic rays or random hits in $Z \to \mu \mu$ are
estimated from the muon track-time distributions ($t_0^f$) using the
same fitting techniques as used for the $W$-background estimate.  The
total contamination from these sources in the final sample is found to
be $(5.1 \pm 3.6) \%.$

\subsubsection{$Z \rightarrow \tau\tau$ Background to $Z \rightarrow ll$}

The process $Z \rightarrow \tau\tau$, where both
taus decay to either electrons or muons is a small
background in this analysis.
In the $Z \rightarrow  ee$ sample, 
the reduced acceptance (due to the soft $p_T$
spectra of the electrons) and small $\tau$ branching fractions
allow us to neglect this background.
For the $Z \to \mu \mu$ sample, we estimate a background of
$(0.7 \pm 0.2) \%.$

\subsubsection{Drell-Yan Pair Production}

The Drell-Yan process $\gamma \rightarrow ll$ is coherent with $Z
\rightarrow ll$, and the experimentally observed production of lepton
pairs corresponds to the square of the sum of the $M_\gamma$ and $M_Z$
amplitudes. However, because we are interested in comparing to
theoretical predictions for the $|M_Z|^2$ term, the size of the
Drell-Yan fraction $|M_{\gamma}|^{2}$ and the interference term must
therefore be deduced before making our comparison.

We use the {\small{ISAJET}} Monte Carlo~\cite{isajet} to estimate
these two terms relative to pure $Z$ production, and express them as
the fraction of the number of pure $Z$ events.  We cross check with
{\small{PYTHIA}}~\cite{pythia} and find a similar number. This
``background'' is thus found to be $(1.2 \pm 0.1) \%$ for the electron
channel, and $(1.7 \pm 0.3) \%$ for the muon channel.

\subsection{Summary of Backgrounds in the $W\to l\nu$ and $Z\to l l$ Samples.}

The backgrounds are summarized in Table~\ref{tab:backs}. We find a
total background in the $W \to l \nu$ samples of $(5.7\pm1.7)\%$ for
electrons and $(22.1\pm1.9)\%$ for muons. For $Z \to l l$, the total
background estimates are $(4.0\pm1.4)\%$ and $(10.1\pm3.7)\%$, for the
electron and muon channels, respectively.

\begin{table}
\begin{center}
\begin{tabular}{c|cccc}
Backgrounds  & $W\to e\nu$ & $W\to\mu\nu$ 
             & $Z\to e e$  & $Z\to\mu\mu$ \\ \hline
QCD Dijets
        & $3.3 \pm 1.7$ & $5.1 \pm 0.8$ 
        & $2.8 \pm 1.4$ & $2.6 \pm 0.8$\\ 
Cosmic Rays, etc.
        & & $3.8 \pm 1.6$ & & $5.1 \pm 3.6$\\ 
$W\to\tau\to e (\mu)$
        & $1.8 \pm 0.2$ & $5.9 \pm 0.5$ & & \\ 
$Z\to ee (\mu\mu)$
        & $0.6 \pm 0.1$ & $6.5 \pm 0.5$ & & \\ 
$Z\to\tau\tau\to\mu(\mu\mu)$
        & & $0.8\pm0.2$   & & $0.7\pm0.2$ \\ 
Drell--Yan  
        & & & $1.2 \pm 0.1$ & $1.7\pm0.3$ \\ \hline 
Total   & $5.7 \pm 1.7$ & $22.1 \pm 1.9$ 
        & $4.0 \pm 1.4$ & $10.1 \pm 3.7$ \\
\end{tabular}
\end{center}
\caption{Backgrounds to the $W$ and $Z$ samples. All estimates
are in per cent of the total candidate samples.}
\label{tab:backs}
\end{table}

\section{Detector Simulation and Acceptance}

The acceptances for the processes $p\bar{p} \rightarrow W \rightarrow
l\nu$ and $p\bar{p} \rightarrow Z \rightarrow l l$ are defined as the
fractions of all the $W \rightarrow l\nu$ or $Z \rightarrow l l$ events
that pass our fiducial and kinematic criteria.  The acceptance is
estimated using an event generator to model vector-boson production and
decay, and a Monte Carlo simulation of the D\O\ detector.  This section
describes the event generators, the detector simulations, and the
results of the acceptance calculations.

\subsection{$W \to e \nu$ and $Z \to e e$ Simulation}

A fast Monte Carlo simulation is used to calculate the acceptance in
the electron channel.
A parton-level generator produces
a vector boson, which is made to decay to leptons in the boson
rest frame. The leptons are boosted to the lab frame 
according to the longitudinal and transverse momentum of the
boson.  The longitudinal momentum of the boson is
determined by the parton distribution functions and the 
energy in the center of mass. The transverse motion
is caused by radiation of initial-state partons or through higher 
order contributions to $W$ or $Z$ production.
We calculate the $p_T$ spectra from the double differential
cross section $d^{2}\sigma/dp_Td\eta$
provided in a next-to-leading order (NLO) program~\cite{Arnold}. 
The calculation uses a standard perturbative method for 
high $p_T$, a resummation scheme for the low $p_T$ region,
and a matching scheme between the two.

The specific procedure involves, first, generation of the rapidity of
the vector boson from the randomly-selected momenta of the incident
quarks. Then, the double differential cross section at that rapidity
value is used to generate a $p_T$ distribution, from which the $p_T$ of
the vector boson is chosen.  Once all the four-vectors of the $W$ or $Z$
boson and decay leptons are generated, the differential cross section is
calculated and used as a weight for this event. One million such events
were generated, and the weights used to obtain the acceptance of our
geometrical and kinematical criteria.

The detector simulation includes modeling of the primary vertex
distributions, the electron energy and
{\hbox{$\rlap{\kern0.25em/}E_T$}}\ resolutions, and the turn-on of the
Level 2 trigger.  The vertex ($z$-position) of collisions is generated
from a Gaussian distribution with $\sigma = 30$ cm and $<z> = -8$ cm, to
reproduce the measured vertex distribution in the data.  The electron
energy is smeared with a resolution of
\begin{equation} \left( \frac{\sigma}{E} \right)^{2} = C^{2} +
\left( \frac{S}{\sqrt{E}} \right)^{2} \end{equation}
where $S=15.7\% \sqrt{{\rm GeV}}$ and C is 0.4\%.
 
The missing $E_T$ is reconstructed from $
{\hbox{$\rlap{\kern0.25em/}E_T$}} = p^{W}_{T} - p^{e}_{T} $, where
$p_T^e$ is smeared according to the EM energy resolution, and the $p_T$
of the $W$ is smeared to match the hadronic energy resolution, because
$p_T^W$ is determined from the hadrons recoiling against a $W$ boson.  A
correction factor of 0.83 is applied to the hadronic energy scale of the
calorimeter; this factor is obtained by studying the balance of the sum
of the $p_T$ of two electrons and the recoil hadrons in $Z$ events.

The effect of underlying events in the data is included in the detector
simulation.  A vector of {\hbox{$\rlap{\kern0.25em/}E_T$}}, chosen at
random from a sample of minimum bias events, is added to the above
{\hbox{$\rlap{\kern0.25em/}E_T$}}\ to simulate the smearing contributed
by the underlying event.

The event simulation includes radiative corrections.  We first calculate
the acceptance for the radiative final states $W\gamma$ or $Z\gamma$
(with a threshold energy for the photon of greater than 20
MeV)~\cite{BerendsKleiss}.  In this calculation, we define a cone whose
axis is centered along the direction of the electron, with the cone size
defined as ${\cal R}=\sqrt{\delta \eta^{2} + \delta \phi^{2}}$, where
$\delta \eta$ and $\delta \phi$ are the differences in $\eta$ and $\phi$
between the electron and photon directions.  If $R < 0.3$, we add the
energies of the electron and photon, and treat the sum as the electron
energy.  Otherwise the energy of the electron is left intact.  We then
combine the acceptances of $W/Z$ and $W\gamma/Z\gamma$ to get the final
acceptances. The radiative corrections are $0.6\%$ for our $W$
measurement and $1.6\%$ for the $Z$ channel.

\begin{figure}[ht]
\centerline{\epsfysize=10cm\epsffile{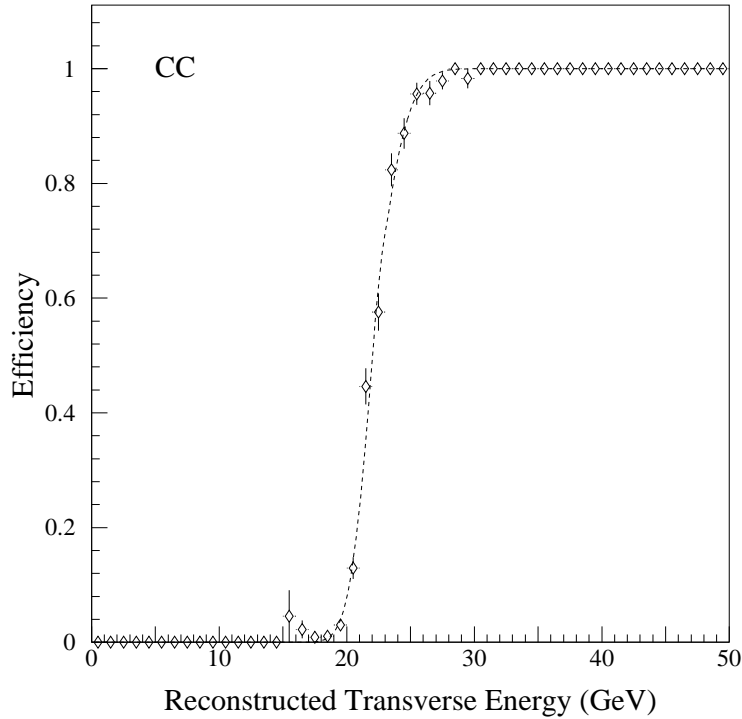}}
\caption{Trigger efficiency  for electrons in the CC
as a function of reconstructed $E_T$.
The threshold behavior is similar for electrons in the EC.
The line is a function fitted to the data points.}
\label{fig:el_turnon}
\end{figure}

The final effect included in the simulation
is the electron $p_T$ cutoff at Level 2.
The effect of this cutoff is folded in using the trigger turn-on curves
observed in the data (see Fig.~\ref{fig:el_turnon}).

\subsection{$W \to \mu \nu$ and $Z \to \mu \mu$ Simulation}

A full detector simulation is used for the acceptance calculations in
the muon channel.  We use {\small{ISAJET}}~\cite{isajet} as the
parton-level event generator, and {\small{GEANT}}~\cite{geant} to model
the D\O\ detector.  We incorporate the PDT efficiencies and resolutions
on a chamber-by-chamber basis using measurements from the data.

The chamber efficiencies are obtained from large samples of good quality
muons, and include variations observed during the run.  We check these
efficiencies by comparing the Level 1 trigger efficiencies predicted
from the simulation with those observed in the data.  The individual
chamber resolutions are obtained from residuals in fits to muon tracks
in collider data.  Finally, the overall momentum resolution in the Monte
Carlo is tuned to fit the shape of the reconstructed $Z \to \mu \mu$ and
$W \to \mu \nu$ mass distributions.

\subsection{Acceptance Calculation}

  We define the acceptance as the fraction of events passing our
kinematic and fiducial requirements. For $Z \to e e$, we also include
the effect of requiring the invariant mass of the lepton pairs to be
within the mass window from 75~GeV to 105~GeV.
This mass requirement accepts $(95.5\pm0.3)\%$ of the events.  For $Z
\to \mu \mu$, we also include the effect of the requirement on the 
angles between the two muons. This has an efficiency of only $(81\pm1)\%$, but
greatly reduces the cosmic ray background.

The overall acceptances $A^W$ for $W \rightarrow l \nu$, and $A^Z$ for
$Z \rightarrow ll$, are summarized in Table~\ref{tab:wzacc}.  The
acceptance in the muon channel is much lower than for the electron
channel because, for this analysis, we restrict the $\eta$ range 
to the region where the single muon trigger is most efficient.
The effect is roughly a factor of two per lepton.

\begin{table}
\begin{center}
\begin{tabular}{l|cc}
            &   $A^{W}$          &    $A^{Z}$  \\ \hline
 Electron Channel &  $(46.0 \pm 0.6)\%$ & $(36.3 \pm 0.4)\%$ \\ 
 Muon Channel     &  $(24.8 \pm 0.7)\%$ & $(6.5  \pm 0.5)\%$  
\end{tabular}
\end{center}
\caption{$W$ and $Z$ acceptances in the electron and muon decay channels.
The systematic errors come from a variety of sources, as explained in 
subsection $V.D$.}
\label{tab:wzacc}
\end{table}

\subsection{Systematic Errors on the Acceptances}
\label{sec:accsys}

We summarize the systematic uncertainties for acceptances of $W$ and $Z$
bosons in the electron channel, and for the ratio of acceptances for the
two processes, in Table~\ref{tab:elacc_sys}.  The errors on the ratio
are calculated separately, taking account of the partial cancelation of
the systematic errors in $A^{W}$ and $A^{Z}$.

The systematics from the parton distribution functions (pdf)
are studied by comparing results obtained with  CTEQ2M, CTEQ2MS,
GRV, MRSD0$^{\prime}$ and MRSD--. We define CTEQ2M as the central value
of our
calculations, recalculate the acceptances for 
each pdf, and quote the maximum difference in our prediction  
as the systematic error. 
We also vary the parameterization of the $p_T$ spectrum used to
generate the $W$ and $Z$ events within the range consistent with our data.

The values of the $W$ mass and width~\cite{PDG} are varied by one
standard deviation and the consequences propagated through the
acceptance calculations.  The $W$ acceptance is sensitive to the mass of
the $W$ boson, varying by $0.7\%$ for a change of mass of $0.18$
GeV. The result is not sensitive to the width of the $W$
boson.  The errors from uncertainties in the mass or width of the $Z$
are extremely small, and we therefore neglect them.  The error in the
simulation of the {\hbox{$\rlap{\kern0.25em/}E_T$}}\ depends mainly on
the smearing and simulation of the energies of soft jets.  This error is
dominated by the uncertainty in the hadronic energy scale.

We vary the parameterizations of the Level-2 trigger efficiency and the
input-vertex distribution to estimate the uncertainties from these
sources.

The electromagnetic energy scale error and resolution also contribute to
the systematic uncertainty on the acceptances.  We vary both of these
within their measured uncertainties, and find that the uncertainties
from both these sources are small.

The error in the acceptances due to radiative corrections
is obtained by varying the cone size used to estimate the corrections.

\begin{table}
\begin{center}
\begin{tabular}{l|ccc}
                      &   $A^{W \rightarrow e\nu}$  
                      &  $A^{Z \rightarrow ee}$  
                      & $A^{W \rightarrow e\nu}/A^{Z \rightarrow ee}$ \\ \hline
 Choice of pdf's      &   $0.4\%$  &  $0.6\%$  & $0.3\%$ \\ 
 $W/Z$ $p_T$ spectra  & $0.3\%$  &  $0.2\%$  & $0.4\%$ \\ 
 $W$ mass             &   $0.7\%$  &   --       & $0.7\%$ \\ 
 $W$ width            &   $<0.2\%$ &   --       & $<0.2\%$ \\
 {\hbox{$\rlap{\kern0.25em/}E_T$}}                &   $0.6\%$  &   --       & $0.6\%$ \\  
 Trigger efficiency   &   $0.3\%$  &  $<0.1\%$ & $0.3\%$  \\ 
 Vertex distribution  &   $0.4\%$  &  $0.4\%$  & $0.3\%$ \\
 EM energy resolution &   $0.1\%$  &  $0.3\%$  & $0.2\%$ \\
 EM energy  scale     &   $0.3\%$  &  $0.3\%$  & $0.2\%$\\ 
 Radiative corrections & $0.3\%$  &  $0.4\%$  & $0.4\%$ \\ 
 $Z$ mass window      &    -       &  $0.3\%$  & $0.3\%$ \\ \hline
 Total                &   $1.3\%$  &  $1.0\%$  & $1.3\%$ \\ 
 \end{tabular}
\caption{Relative uncertainties in the acceptances for $W\to e \nu$
and $Z\to ee$ events.  Details of these estimates are given in the text.}
\label{tab:elacc_sys}
\end{center}
\end{table}

The muon channels have quite different, and, in general, larger
systematic uncertainties than the electron channels. 
These uncertainties are summarized in  Table~\ref{tab:muacc_sys}.

We determine the systematic error from 
the chamber efficiencies by varying the efficiencies within their
uncertainties and repeating the
acceptance calculation. 
We determine the uncertainty in resolution by varying the overall
resolution within the constraint that it be consistent with 
the observed $Z \to \mu \mu$ mass distribution.

\begin{figure}[ht]
\centerline{\epsfysize=10cm\epsffile{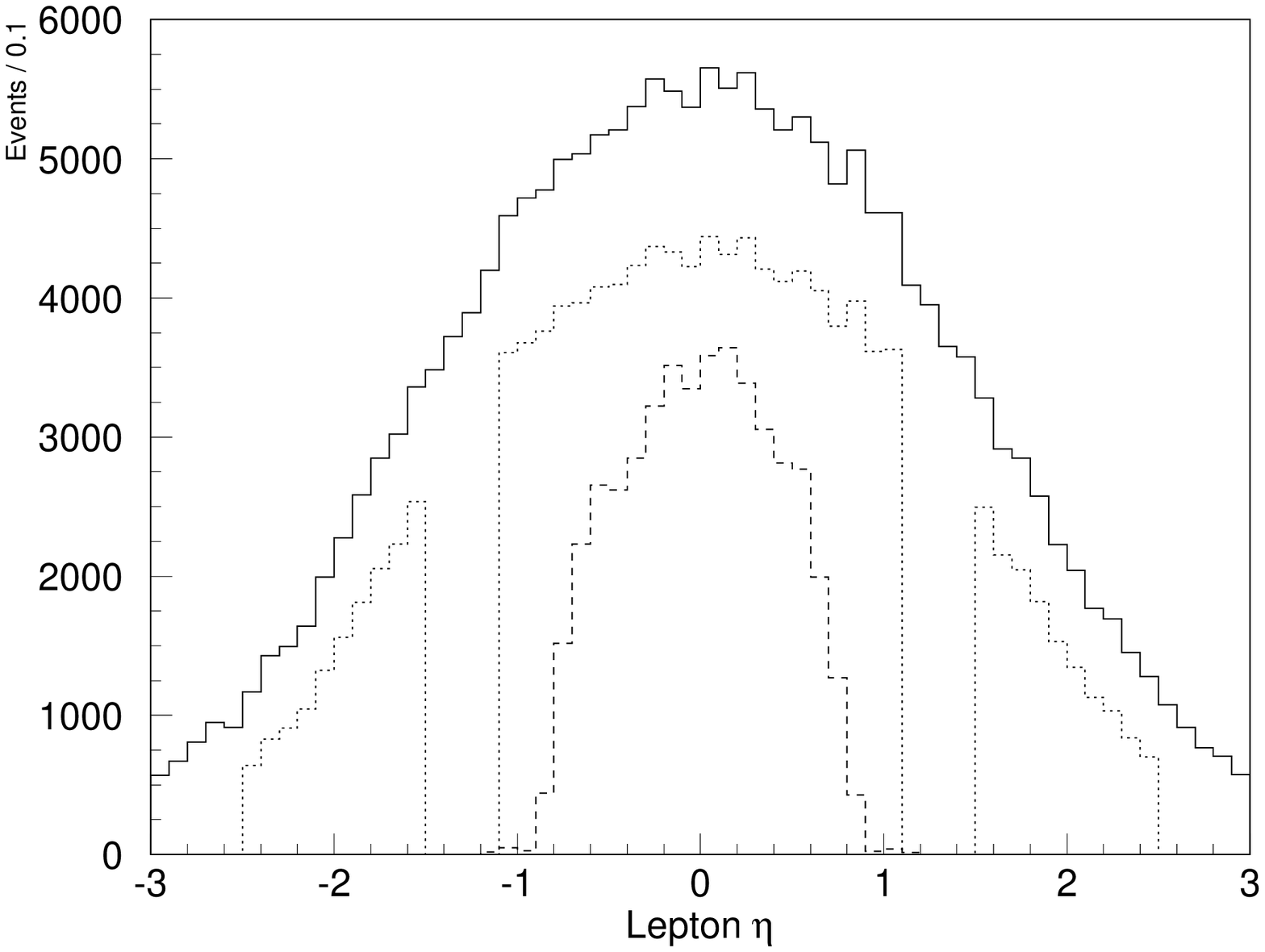}}
\caption{Pseudorapidity distributions for charged leptons in $W\to l\nu$ events.
The solid line is the generated distribution for either electrons or muons.
The dotted and dashed lines show the distributions for the electron
and muon channels, respectively, after applying the fiducial requirements.}
\label{fig:acceta}
\end{figure}

The dependence on pdf is larger for the muon channel than for the
electron channel because of the difference in $\eta$ coverage.  The muon
rapidity range ends in a region of large acceptance, and so is quite
sensitive to the input parton distributions, while the electron coverage
is more complete, extending in pseudorapidity to a region where the 
contribution to the cross section
is relatively small, and so is less sensitive to the
inputs.  Figure~\ref{fig:acceta} shows the generated pseudorapidity
distribution for $W\to\ell\nu$ and accepted regions for muons and
electrons.

\begin{table}
\begin{center}
\begin{tabular}{l|ccc}
            &   $A^{W \rightarrow \mu\nu}$          
         &    $A^{Z \rightarrow \mu\mu}$  &  
        $A^{W \rightarrow \mu\nu}/A^{Z \rightarrow \mu\mu}$    \\ \hline
 Choice of pdf's        & 2.0\% & 4.0\% & 2.0\% \\
 Chamber efficiencies   & 0.7\% & 4.4\% & 3.6\%\\
 Chamber resolutions    & 1.0\% & 0.4\% & 1.4\% \\
 Monte Carlo statistics & 1.7\% & 3.9\% & 4.2\% \\ \hline
 Total & 2.9\% & 7.1\% & 6.0\%  \\
 \end{tabular}
\caption{Systematic uncertainties in the acceptance for $W\to\mu\nu$
and $Z\to\mu\mu$ events. All other systematic uncertainties are negligible
compared to those listed in this table.}
\label{tab:muacc_sys}
\end{center}
\end{table}

\section{Calculations of Efficiency}

To measure the efficiency for particle identification (ID) and
triggering requires choosing clean and unbiased samples of electrons
and muons.  We use the $Z \to e e$ and $Z \to \mu \mu$ samples as our
source of high-$p_T$ leptons with low background.  Requiring only one
lepton to pass all the particle ID and trigger criteria, leaves the
other lepton unbiased with respect to these cutoffs, and it can be
used to measure the efficiencies.

\subsection {Electron Efficiencies}

For obtaining electron efficiencies, we use the $Z \to e e$ events in
the peak region $86 < M_{ee} < 96$ GeV.  As usual, we determine
the systematic error by varying the electron selection criteria and
the procedure used for background subtraction.  We used two sets of
selection cuts, one being the tight cutoffs, and one being the
standard cutoffs (see Sect III.A.6).

The major background in the $Z$ sample is from QCD jet production with
two jets misidentified as electrons. The two-body invariant mass
distributions are used to study this background.  As discussed
earlier, over a limited mass range, this background can be
approximated by an exponential dependence, or by simple polynomials.
We extract the background fraction in three ways:
\begin{itemize}
\item Method 1:
The average of the number of events in the two sideband regions 
of the $Z$ peak, 
$61 < M_{ee} < 71$ GeV and $111 < M_{ee} < 121$ GeV,  
is taken as the background in the peak region. 
  
\item Method 2:
The mass spectrum is fitted to the sum of 
a relativistic Breit-Wigner shape for the $Z$
convoluted with a Gaussian resolution for $M_{ee}$, 
and a linear function for the background. 
The result of the fit in the region 
$60 < M_{ee} < 70 $ GeV is used to estimate the background.
 
\item Method 3: 
The same fit in Method 2 is used but the two sideband regions
$M_{ee} <$ 70 GeV and $M_{ee} >$ 110 GeV
are used to measure the background under the $Z$ peak.

\end{itemize}

The efficiency of a particular set of criteria is measured by
\begin{equation} \varepsilon = \frac{\varepsilon_{s} - \varepsilon_{b}f_{b}}
{1-f_{b}} \end{equation} 
where $\varepsilon_{s}$ is the efficiency measured for 
events in the peak region, $\varepsilon_{b}$ is 
that measured for events in the background region and $f_{b}$ is the 
fraction of background in the peak region. Note that it does not matter
that there are some real $Z \to e e$ events in the background region.
Our measurement of the efficiencies is correct as long as there is
only signal left in the peak region after background subtraction.

\subsubsection{Efficiency for Electron Selection}

The single-electron selection efficiency can be expressed as 
\begin{equation} \varepsilon = 
\varepsilon_{e\_\rm{shwr}}\cdot\varepsilon_{e\_\rm{trk}}
\cdot\varepsilon_{e\_\rm{trig}},\end{equation}
where $\varepsilon_{e\_\rm{shwr}}$ is the efficiency from the
requirements on shower shape in offline reconstruction,
$\varepsilon_{e\_\rm{trk}}$ is the efficiency of associating a good
track with an electron cluster in the calorimeter, and
$\varepsilon_{e\_\rm{trig}}$ is the efficiency of the trigger.

The $\varepsilon_{e\_\rm{shwr}}$ values reflect the following selections
\begin{itemize}
    \item H-Matrix $\chi^{2} < 100$  
    \item $f_{\rm{iso}} < 0.10$
    \item $f_{\rm{EM}} > 0.95$
\end{itemize}
Efficiencies for electrons are measured separately in the CC and in the
EC.  We extract six values, corresponding to the two sets of tagging
criteria (standard and tight), and the three background estimations
discussed above.  The central value is taken to be the median value
closest to the mean, and its uncertainty as the combination of the
statistical uncertainty associated with the central value and half the
difference between the highest and lowest calculated efficiencies.  The
results are
\begin{eqnarray*}
 \varepsilon_{\rm{shwr\_CC}} &=& 0.881\pm0.015
\end{eqnarray*}
and
\begin{eqnarray*}
 \varepsilon_{\rm{shwr\_EC}} &=& 0.884\pm0.024, 
\end{eqnarray*}
where the errors are dominated by statistics.

The efficiency for reconstructing a track associated with an electron 
has two
components: the efficiency for finding a track near the electron, and
the efficiency of the cutoff on the parameter $\sigma_{\rm{trk}}$, which
provides the quality of the match between the position of the track the
calorimeter shower.  The efficiency of track matching is measured by
taking the ratio of the number of calorimeter clusters in the $Z \to e
e$\ sample that are reconstructed as electrons to the number
reconstructed as either electrons or photons.  (The only difference is
the presence or absence of a track match.)

The $\sigma_{\rm{trk}}$ and trigger efficiencies are obtained in the
same way as the efficiencies for shower shape cuts in the calorimeter.
Combining $\varepsilon_{e\_\rm{trk}}$ and $\varepsilon_{e\_\rm{trig}}$,
using our six estimations, we obtain
\begin{eqnarray*}
 \varepsilon_{\rm{trk\_CC}}\cdot \varepsilon_{\rm{trig\_CC}} &=& 0.830\pm0.014
\end{eqnarray*}
and
\begin{eqnarray*}
 \varepsilon_{\rm{trk\_EC}}\cdot \varepsilon_{\rm{trig\_EC}} &=& 
0.774\pm0.024. 
\end{eqnarray*}
The uncertainties are again dominated by the statistics
of the $Z$ event sample.

\subsubsection{Efficiencies for $W \to e \nu$ 
and $Z \to e e$ Events}

The selection criteria for electrons from $W$ events are just the ones
given above (tight). The total selection efficiency is therefore the
product of the shower, track, and trigger terms, giving
\begin{equation} \varepsilon^{W \to e \nu}_{\rm{CC}} = 
0.731\pm0.018\end{equation}
and 
\begin{equation} \varepsilon^{W \to e \nu}_{\rm{EC}} = 
0.684\pm0.028.\end{equation} 
We combine the CC and EC results, by weighting them by their relative 
acceptances, to obtain
\begin{equation} \varepsilon^{W \to e \nu} =  0.704 \pm 0.017.\end{equation}
 
For $Z \to e e$ events, one of the two electrons is selected exactly as
in $W \to e \nu$ events.  The second one is selected without imposing
any track or trigger requirements, but with all the other criteria being
the same; thus the total efficiency for selecting this loose electron is
just $\varepsilon_{\rm{shwr}}$.

The $Z \to e e$\ efficiencies are therefore
\begin{eqnarray*}
 \varepsilon^{Z \to ee}_{\rm{CC-CC}} &=& 0.753\pm0.020, \\
 \varepsilon^{Z \to ee}_{\rm{CC-EC}} &=& 0.748\pm0.023,
\end{eqnarray*}
and
\begin{eqnarray*}
 \varepsilon^{Z \to ee}_{\rm{EC-EC}} &=& 0.742\pm0.042. 
\end{eqnarray*}
The overall efficiency is obtained by weighting using the relative
acceptances of CC and EC events, giving
\begin{equation} \varepsilon^{Z \to ee} = 0.736 \pm 0.024.\end{equation}
  
The ratio of efficiencies of $W$ to $Z$ selections is calculated
directly using each of our different methods, and the systematic error
assigned independently from $\varepsilon^{W}$ or $\varepsilon^{Z}$. In
this way, any correlation of systematic errors of $\varepsilon^{W \to e
\nu}$ and $\varepsilon^{Z \to ee}$ is taken into account.  The ratio of
efficiencies is
\begin{equation} \frac{\varepsilon^{Z \to ee}}{\varepsilon^{W \to e \nu}} 
= 1.045\pm0.019.\end{equation}

\subsection{Muon Efficiencies}

The single-muon efficiency can be written as
\begin{equation} \varepsilon = \varepsilon_{\mu\_\rm{reco}} 
\cdot \varepsilon_{\mu\_\rm{ID}}
\cdot \varepsilon_{\mu\_\rm{trig}} \end{equation}
where $ \varepsilon_{\mu\_\rm{reco}}$ is the muon reconstruction efficiency,
$ \varepsilon_{\mu\_\rm{ID}}$ is the muon ID efficiency, and
$ \varepsilon_{\mu\_\rm{trig}}$ is the muon trigger efficiency. 
Each of these efficiencies is measured using a different unbiased
data sample, as described below.

\subsubsection{Muon Reconstruction Efficiency}

The muon reconstruction efficiency is estimated using events from a
special data run that had no Level-2 requirements.  We require that
there be a jet reconstructed offline in the same $\eta$-$\phi$ region as
a muon candidate found by the Level-1 trigger.  No Level-2 or muon
reconstruction criteria are imposed either online or offline, and the
muon candidates are categorized as ``good'' or ``bad'' tracks through
visual examination of the event displays.  The efficiency is defined to
be the percentage of ``good'' muon tracks which pass the loose muon ID
(defined in Sec.~\ref{sec:muid}).  
To reduce the systematic error of this method,
the scanning is performed by at least two physicists.  The offline
reconstruction efficiency for ``good'' tracks is found to be
\begin{equation} \varepsilon_{\mu\_\rm{reco}} = 0.952\pm0.033 \end{equation}
per muon.

\subsubsection{Muon Identification Efficiency}

Muon efficiencies are derived directly from the $Z \to \mu \mu$ sample.
One muon tags the event by passing all the particle ID requirements,
leaving the other muon to form part of an unbiased sample of isolated
high-$p_T$ muons.  The $p_T$ cutoff is raised to 20 GeV on both
muons, which minimizes the backgrounds from QCD, cosmic rays, and random
PDT hits.

The overall muon ID efficiency is found to be
\begin{equation} \varepsilon_{\mu\_\rm{ID}} = 0.626 \pm 0.047. \end{equation}
The most important contributions to the inefficiency stem
from the requirement on track matching in the CD 
($\varepsilon = 0.82\pm0.04)$,
for finding a track and matching the angles and from the
calorimeter isolation criteria ($\varepsilon = 0.85\pm0.04$).

\subsubsection{Muon Trigger Efficiency}

The muon trigger efficiency is not estimated from the $Z \to \mu \mu$
sample because of poor statistics. Instead, a sample of ``unbiased''
muons that pass quality criteria, and are present in events passing a
non-muon (usually a jet) trigger, are used.  The results are
cross--checked with the ones obtained from the $Z \to \mu \mu$ sample,
and the results agree within the statistical uncertainty.

The overall trigger efficiency for high-$p_T$ muons is
\begin{equation} \varepsilon_{\mu\_\rm{trig}} = 0.367 \pm 0.019. \end{equation}
The relatively limited geometric coverage of the muon chambers is the
most important factor contributing to this low efficiency.  The Level-1
trigger requires hits in all three layers of the PDT system, and only
$\approx 60\%$ of muon tracks in the fiducial region satisfy this
requirement. For tracks that satisfy it, the trigger efficiencies are
$(80\pm2)\%$ for Level 1 and $(78\pm3)\%$ for Level 2.

\subsubsection{$W \to \mu \nu$ and $Z \to \mu \mu$ Efficiencies}

The efficiency for $W \to \mu \nu$ candidates corresponds to the
single-muon efficiency described above.  The total efficiency is
therefore the product of the reconstruction, ID, and trigger terms, and
equals
\begin{equation} \varepsilon^{W \to \mu \nu} =  0.219 \pm 0.026.\end{equation}
 
The $Z \to \mu \mu$\ efficiency takes into account the fact that both
muons must be reconstructed, but that only one has to pass the ID and
trigger criteria.  Combining the single muon efficiencies, we obtain
\begin{equation} \varepsilon^{Z \to \mu \mu} = 0.527 \pm 0.049.\end{equation}
  
The ratio of efficiencies of $W$ to $Z$ selections takes account of the
correlations among the systematic errors. The result is
\begin{equation} \frac{\varepsilon^{Z \to \mu \mu}}
{\varepsilon^{W \to \mu \nu}} = 2.48\pm0.19. \end{equation}

\section{Cross Section}
  
\subsection{Determination of Integrated Luminosity}

The integrated luminosity is determined by monitoring non-diffractive
inelastic {\mbox{$p\bar p$}} collisions using two hodoscopes of
scintillation counters (the Level-0 trigger~\cite{D0det}) mounted on the
front surfaces of the end calorimeters near the beam axis.  The
average~\cite{d0_lum} of the values measured by the CDF~\cite{cdf_lum}
and E710~\cite{e710_lum} experiments at Fermilab is used for the
inelastic cross section.  The reaction rate measured by the Level-0
system corresponds to a cross section (the Level-0 visible cross
section) of $ \sigma_{\rm{L0}} = 46.7$ mb.

For the electron trigger used in this analysis, after corrections for 
experimental dead times and multiple interactions,
the integrated luminosity is determined
to be 
\begin{equation} \int Ldt = 12.8 \pm 0.7 \mbox{ pb}^{-1},\end{equation}
while the muon trigger had an exposure of
\begin{equation} \int Ldt = 11.4 \pm 0.6 \mbox{ pb}^{-1}.\end{equation}
The 5.4\% systematic uncertainty in the luminosities is
calculated from the uncertainty in
the $p\bar{p}$ inelastic cross section (4.6\%), 
the systematic errors on the acceptance (2.0\%), and efficiency (2.0\%)
of the Level-0 detectors. 

\subsection{$W$ and $Z$ Production Cross Sections}

\subsubsection{Theoretical Predictions}

The $W$ and $Z$ boson total production cross sections have been computed
from a complete calculation to order $\alpha_s^2$~\cite{HVM}.
We used $M_Z = 91.19$ GeV, $M_W = 80.23$ GeV,
and $\sin^2 \theta_w = 1 - M_W^2 / M_Z^2 = 0.226$,
and CTEQ2M~\cite{cteq} pdf for our central value 
and considered the other pdf sets shown in Table~\ref{tab:xsec}.
The strong correlation of the $W$ and $Z$ boson cross sections 
decreases the sensitivity of the ratio of cross sections to variations
of the pdf.
Taking CTEQ2MS and CTEQ2ML as the extremes,
we obtain $\sigma_W / \sigma_Z = 3.33 \pm 0.02.$

\begin{table}
\begin{center}
\begin{tabular}{c|ccc}
pdf             & $\sigma_W$ (nb) & $\sigma_Z$  (nb) 
& $\sigma_W/\sigma_Z$ \\ \hline
MRSS0$\prime$   & 22.114 &  6.633  &    3.334    \\ 
MRSD0$\prime$   & 22.150 &  6.680  &    3.316    \\ 
MRSD$\prime$-   & 21.810 &  6.558  &    3.326    \\ 
MRSH            & 22.043 &  6.594  &    3.343    \\ 
MRSA            & 22.054 &  6.651  &    3.316    \\ 
CTEQ2M          & 22.350 &  6.708  &    3.332    \\ 
CTEQ2MS         & 21.662 &  6.541  &    3.312    \\ 
CTEQ2MF         & 22.589 &  6.788  &    3.328    \\ 
CTEQ2ML         & 23.357 &  6.963  &    3.354    \\ \hline
MRSD$\prime-$ DIS & 22.190 &  6.662  &    3.331    \\ 
MRSH DIS        & 22.404 &  6.691  &    3.348    \\ 
CTEQ2D          & 22.670 &  6.719  &    3.374    \\
\end{tabular}
\end{center}
\caption{$W$ and $Z$ boson production cross section predictions, calculated
using different pdf sets.
All sets are in the $\overline{MS}$, scheme 
except the last three, which use the DIS scheme.}
\label{tab:xsec}
\end{table}

Until recently, the uncertainties on the calculated cross sections were
dominated completely by the variation due to choice of pdf.  Recent
measurements of the proton structure function $F_2$ and of the $W^{\pm}$
rapidity distributions have restricted the acceptable pdf choices to the
point that other sources of error must be considered.  The sources we
considered are the use of NLO pdf sets instead of NNLO (which would be
more appropriate for use with the ${\cal O}(\alpha_s^2)$ calculation),
variation of the calculated cross section from the uncertainty in $M_W$,
and the uncertainty due to the dependence on renormalization and
factorization scales.

While the $W$ and $Z$ boson total cross sections have been calculated up
to ${\cal O}(\alpha_s^2)$, the corresponding NNLO pdf sets are not yet
available.  The uncertainty on the cross sections due to using NLO pdf's
has been estimated~\cite{neerven} to be 3\% at $\sqrt{s} = 1.8$
TeV. This uncertainty is assumed to cancel in the predicted ratio of
cross sections.

The error on the mass of the $W$ boson leads to an uncertainty in the
value of the $W$ cross section, and to a lesser extent the $Z$ cross
section (since $\sin^2 \theta_w$ is correlated with $M_W$).  The effect
on the individual cross sections is small compared to that from the
choice of pdf.  However, in the ratio of cross sections, the two
contributions are comparable.  The effect of the $M_W$ uncertainty is
shown in Table~\ref{tab:mw}.

\begin{table}
\begin{center}
\begin{tabular}{c|ccc}
$M_W$ (GeV) & $\sigma_W$ (nb) & $\sigma_Z$ (nb) & $\sigma_W/\sigma_Z$ \\
\hline
80.05 & 22.403 &  6.671   &  3.358  \\ 
80.23 & 22.350 &  6.708   &  3.332  \\ 
80.41 & 22.298 &  6.745   &  3.306  \\
\end{tabular}
\end{center}
\caption{$W$ and $Z$ boson production cross section predictions, 
calculated for
values of $M_W$ one standard deviation below and above the world average
(using the CTEQ2M
pdf).}
\label{tab:mw}
\end{table}

The last source of error considered for the variation of the calculated
cross sections is the choice of factorization and renormalization
scales.  It is customary to set both scales equal to the same
value~\cite{neerven}.  We set the scales to the corresponding values of
the vector boson masses.  The uncertainty is estimated by varying the
scales by a factor of two in either direction.  The results are shown in
Table~\ref{tab:scale}.  The effect is small for the individual cross
sections, as well as for the ratio.

\begin{table}
\begin{center}
\begin{tabular}{ c|ccc}
Scale (GeV) & $\sigma_W$ (nb) & $\sigma_Z$ (nb) & $\sigma_W/\sigma_Z$ \\
\hline
$M_V/2$ & 22.259 &  6.688    &  3.328 \\ 
$M_V$   & 22.350 &  6.708    &  3.332 \\ 
$2M_V$  & 22.421 &  6.715    &  3.339 \\
\end{tabular}
\end{center}
\caption{Calculations of $W$ and $Z$ boson production cross sections for
different values of the factorization and renormalization scales.  
$M_V$ is the mass of the 
corresponding vector boson.  
(The CTEQ2M pdf and the nominal value of  $M_W$ are used.)}
\label{tab:scale}
\end{table}

The effects of all the sources of error on the calculated cross sections 
are summarized in Table~\ref{tab:errors}.  Using CTEQ2M for the
central values, the theoretical predictions for the production cross sections
at $\sqrt{s} = 1.8$ TeV are:

\begin{eqnarray*}
\sigma_W \equiv \sigma(\overline{p}p \rightarrow W + X) & = &
                                    22.4 ^{+1.2}_{-1.0}\ {\rm nb,} \\
\sigma_Z \equiv \sigma(\overline{p}p \rightarrow Z + X) & = &
                                    6.71 ^{+0.33}_{-0.27}\ {\rm nb,} 
\end{eqnarray*}
\begin{eqnarray*}
\frac{\sigma_W}{\sigma_Z} &=& 3.33 \pm 0.03.
\end{eqnarray*}

\begin{table}
\begin{center}
\begin{tabular}{c|ccc}
Error Source & $\delta\sigma_W$ (nb) & $\delta\sigma_Z$ (nb) & 
$\delta(\sigma_W/\sigma_Z)$ \\ \hline
pdf choice   & +1.007, $-0.688$ & +0.255, $-0.167$ & +0.022, $-0.020$ \\
NLO pdf's    & +0.671, $-0.671$ & +0.201, $-0.201$ & --- \\
$\delta M_W$ & +0.053, $-0.052$ & +0.037, $-0.037$ & +0.026, $-0.026$ \\
Scale        & +0.071, $-0.091$ & +0.007, $-0.020$ & +0.007, $-0.004$ \\ \hline
Total Error  & +1.213, $-0.967$ & +0.327, $-0.265$ & +0.034, $-0.033$ \\ 
\end{tabular}
\end{center}
\caption{Summary of estimated uncertainties on the calculated $W$ and 
$Z$ boson production cross sections. The separate errors are added in 
quadrature to form the total error
(assuming no correlation between error sources).}
\label{tab:errors}
\end{table}

In order to compare the theoretical predictions of the $W$ and $Z$ boson
production cross sections to experiment, it is necessary to multiply the
cross sections by the vector boson branching fractions into the
observed experimental channels.  For the current study, only the
electron and muon channels are used.

Very precise values are available for the $Z$ leptonic branching
fractions.  LEP measurements give~\cite{PDG} B$(Z\to\ell\ell) = (3.367
\pm 0.006)\%$.  For the $W$ boson, we use a higher-order theoretical
calculation~\cite{rosner} B$(W\to\ell\nu) = (10.84 \pm 0.02)\%$.
Combining these branching fractions with the production cross sections
quoted above gives the following predicted values for the cross section
times branching fraction:

\begin{eqnarray*}
\sigma_W {\rm{B}}(W \rightarrow \ell\nu)  & = & 
                                    2.42 ^{+0.13}_{-0.11}\ {\rm nb,} \\
\sigma_Z {\rm{B}}(Z \rightarrow \ell\ell) & = & 
                                    0.226 ^{+0.011}_{-0.009}\ {\rm nb.} \\
\end{eqnarray*}

\subsubsection{Results from Experiment}

The cross sections measured for $W$ and $Z$ boson production are calculated 
using the following formula:
\begin{eqnarray*}\sigma \rm{B} =
\frac{N_{\rm{obs}}(1-f_{\rm{bgd}})}{A \varepsilon \int Ldt}. \end{eqnarray*}
where $\rm{N_{obs}}$ is the number of events in our final data sample,
$\rm{f_{bgd}}$ is the fraction of the sample calculated to arise from
background, $A$ is the acceptance of the detector, $\varepsilon$ is the
efficiency for accepted events to reach the final sample, and $\int \rm{Ldt}$
is the integrated luminosity.

\begin{figure*}
\centerline{\epsfxsize=15cm\epsffile{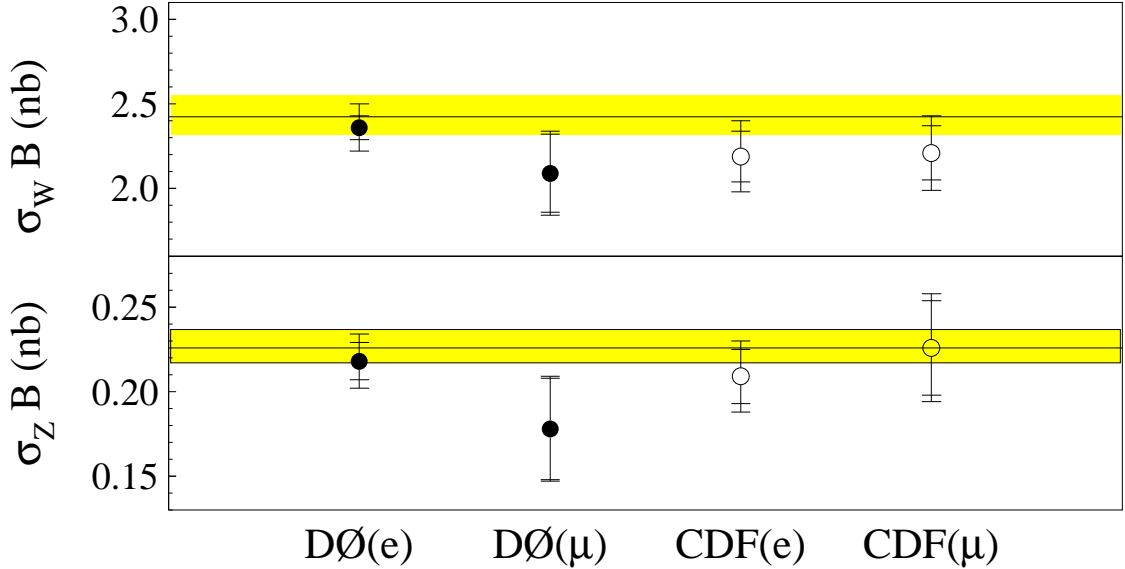}}
\caption{Measurements and predictions
for  $W \to l \nu$ and $Z \to l l$ cross sections.
The results for this experiment are plotted as filled circles
and those for the CDF experiment as open circles.
The inner error bars represent the combined statistical and systematic
uncertainties and the outer error bars include the 
uncertainty on integrated luminosity.
The bands correspond to the range of predictions
discussed in the text.}
\label{fig:xsec_world}
\end{figure*}

The results
are summarized in Table~\ref{tab:crosssection}.
Within the total errors, the measured cross sections are in 
good agreement with theoretical expectations.
Our measurements
are plotted, together with the
predictions and other published experimental
results~\cite{CDF1aX}
at $\sqrt{s} = 1.8$ TeV, in Fig.~\ref{fig:xsec_world}.

\begin{table*}
\begin{tabular}{c|c|c|c|c}
Channel                             & $W\rightarrow{e}\nu$    &
                                      $Z\rightarrow{e^+e^-}$  &
                                      {\mbox{$ W\rightarrow \mu \nu$}}      &
                                      {\mbox{$ Z\rightarrow {\mu^+\mu^-}$}}  \\
\hline
$N_{\rm obs}$                             & $ 10338 $         &
                                            $   775 $         &
                                            $  1665 $         &
                                            $    77 $         \\
Total Bgd(\%)                       & $ 5.7 \pm 1.7 $         &
                                        $ 4.0 \pm 1.4 $         &
                                        $22.1 \pm 1.9 $         &
                                        $10.1 \pm 3.7 $         \\
Acceptance(\%)                      & $46.0 \pm 0.6 $       &
                                      $36.3 \pm 0.4 $       &
                                      $24.8 \pm 0.7 $       &
                                      $ 6.5 \pm 0.4 $       \\
$\epsilon_{\rm trig}\times\epsilon_{\rm sel}$(\%)  
                                    & $70.4 \pm 1.7 $       &
                                      $73.6 \pm 2.4 $       &
                                      $21.9 \pm 2.2 $       &
                                      $52.7 \pm 4.9 $       \\
$\rm \int{{\cal{L}}dt}\ ({\mbox{${\rm pb}^{-1}$}})$  & $ 12.8\pm 0.7  $       &
                                        $ 12.8\pm 0.7  $       &
                                        $ 11.4\pm 0.6  $       &
                                        $ 11.4\pm 0.6  $       \\
\hline
$\sigma$B\enskip (nb)$\pm$(stat),                  &  $ 2.36 \pm 0.02$      &
                                          $ 0.218\pm 0.008   $      &
                                          $ 2.09\pm 0.06     $      &
                                          $ 0.178 \pm 0.022  $      \\
(sys),(lum)  & $\pm 0.08  \pm 0.13   $ \hspace{0.20cm} &
                           $\pm 0.008 \pm 0.012  $ \hspace{0.20cm} &
                           $\pm 0.22  \pm 0.11   $ \hspace{0.20cm} &
                           $\pm 0.021 \pm 0.009  $ \hspace{0.20cm} \\
\end{tabular}
\caption{Observed cross section multiplied by the leptonic branching fraction
for $W$ and $Z$ boson production.}
\label{tab:crosssection}
\end{table*}

\subsection{Extraction of $B(W\to\ell\nu)$ and $\Gamma(W)$ from $R$}   

\subsubsection{Phenomenological Considerations}

The leptonic branching fraction and the total decay width 
of the $W$ boson can be
extracted from the measured ratio of the cross sections 
multiplied by the branching 
fractions of the $W$ and $Z$ bosons into leptons.  
The ratio $R$ can be expressed as follows:

\begin{equation} R \equiv \frac{\sigma_W \rm{B}(W \rightarrow \ell\nu)}
                 {\sigma_Z \rm{B}(Z \rightarrow \ell\ell)} = 
\frac{\sigma_W}{\sigma_Z}\frac{1}{\rm{B}(Z \rightarrow \ell\ell)}
\frac{\Gamma(W \rightarrow \ell\nu)}{\Gamma(W)}. \end{equation}

Using an experimental result for $R$, the known 
$\rm{B}(Z \rightarrow \ell\ell)$, and the prediction of 
${\sigma_W}/{\sigma_Z}$, 
a value for the leptonic
branching fraction of the $W$ boson follows:

\begin{equation} \frac{\Gamma(W \rightarrow \ell\nu)}{\Gamma(W)} = 
\frac{\rm{B}(Z \rightarrow \ell\ell)}{\sigma_W/\sigma_Z} R = 
(0.01011 \pm 0.00011) R. \end{equation}

Alternatively, using, in addition, a calculation of $\Gamma(W
\rightarrow \ell\nu)$, the full width of the $W$ boson width $\Gamma(W)$
can be extracted:

\begin{equation} \Gamma(W) = 
\frac{ (\sigma_W/\sigma_Z)\Gamma(W \rightarrow \ell\nu)}
     {\rm{B}(Z \rightarrow \ell\ell)} \frac{1}{R}. \end{equation}

The leptonic width of the $W$ boson can be written as:
\begin{equation} \Gamma(W \rightarrow \ell\nu) = 
\frac {G_F M_W^3}{6\pi\sqrt{2}}(1 + \delta^{\rm{SM}}). \end{equation}
The corrections $\delta^{\rm{SM}}$ have been calculated in the standard
model by Rosner {\it et al.}~\cite{rosner}.  Using $G_F = (1.16639 \pm
0.00002) \times 10^{-5}$ GeV$^{-2}$, $M_W = 80.23 \pm 0.18$ GeV and
$\delta^{\rm{SM}} = -0.35\%$ gives $\Gamma(W \rightarrow \ell\nu) =
0.2252 \pm 0.0015$ GeV, where the error is entirely due to the
dependence on $M_W$.

In order to properly calculate the uncertainty on $\Gamma(W)$, it is
necessary to take into account the correlation of errors on
$\sigma_W/\sigma_Z$ and $\Gamma(W \rightarrow \ell\nu)$ due to
dependence on $M_W$.  The product of these factors is shown in
Table~\ref{tab:mwerr} for a one standard deviation variation in $M_W$.
Taking the side with the larger variation as the error, the variation in
the product is 0.0009 GeV.  The error on the product due to other
sources is 0.0045 GeV; combining the errors in quadrature gives 0.0046
GeV.  The product, using the nominal value of $\sigma_W/\sigma_Z =
3.33$, is then

\begin{equation} \frac{\sigma_W}{\sigma_Z}\Gamma(W \rightarrow \ell\nu) = 
0.7499 \pm 0.0046\ {\rm GeV}. \end{equation}

Finally, using this value in the expression for $\Gamma(W)$ leads to
\begin{equation} \Gamma(W) = (22.272 \pm 0.137) \frac{1}{R} \rm{GeV}. \end{equation}

\begin{table}
\begin{center}
\begin{tabular}{c|ccc}
$M_W$ &   $\sigma_W/\sigma_Z$ &   $\Gamma(W \rightarrow \ell\nu)$ &
$(\sigma_W/\sigma_Z)\Gamma(W \rightarrow \ell\nu)$ \\ 
(GeV) &  & (GeV) & (GeV) \\ \hline
80.05 &   3.358   &   0.2237 &   0.7512 (+0.0008) \\ 
80.23 &   3.332   &   0.2252 &   0.7504           \\ 
80.41 &   3.306   &   0.2267 &   0.7495 ($-0.0009$) \\
\end{tabular}
\end{center}
\caption{Calculation of the product 
$(\sigma_W/\sigma_Z)\Gamma(W \to \ell\nu)$ 
for a 1 $\sigma$ variation in $M_W$, using the pdf CTEQ2M.}
\label{tab:mwerr}
\end{table}

\subsubsection{Result of Measurements}

\begin{figure}[ht]
\centerline{\epsfysize=10cm\epsffile{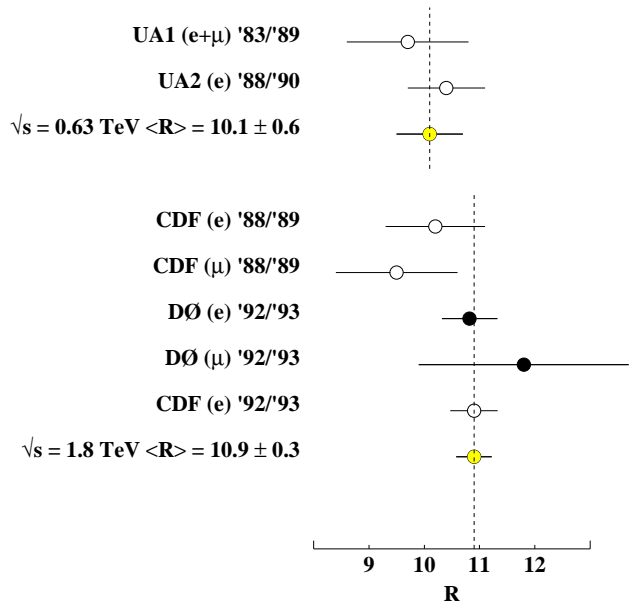}}
\caption{Measurements of the ratio of the
$W \to l \nu$ and $Z \to l l$ cross sections 
multiplied by their respective branching fractions.
The results are shown as a function of the years of the data run.}
\label{fig:r_world}
\end{figure}

The ratio of cross sections is given by: 
\begin{eqnarray*}
R  & = & \frac{N^{W}_{\rm{obs}}(1-f^{W}_{\rm{bgd}})}
{N^{Z}_{\rm{obs}}(1-f^{Z}_{\rm{bgd}})} 
\frac{\varepsilon^{Z}}{\varepsilon^{W}} 
\frac{A^{Z}}{A^{W}}.
\end{eqnarray*}
(The dependence on the luminosity is completely canceled in
the ratio.)
Our results for $e$ and $\mu$ channels are:
\begin{eqnarray*}
 R_{e}     &=& 10.82\pm0.41(\mbox{stat})\pm0.35(\mbox{syst}), \\
 R_{\mu}   &=& 11.8^{+1.8}_{-1.4}(\mbox{stat})\pm1.1(\mbox{syst}),
\end{eqnarray*}
and combined:
\begin{eqnarray*}
 R_{e+\mu} &=& 10.90\pm0.52(\mbox{stat$\oplus$syst}).
\end{eqnarray*}
This is consistent with previous measurements shown in
Fig.~\ref{fig:r_world}.

Using this result, we obtain the branching fraction
\begin{equation} B(W \to l \nu) = (11.02\pm0.52) \%. \end{equation}

Combining this measurement with the calculation of the partial width of
the $W$ boson $\Gamma(W \rightarrow l\nu)$, we obtain
\begin{equation} \Gamma(W) = 2.044\pm0.097 \mbox{ GeV}. \end{equation}
This is in excellent agreement with the prediction of the standard
model, $\Gamma(W) = 2.077\pm0.014$ GeV~\cite{rosner,wmass}, and with the
world average value, $\Gamma(W) = 2.06\pm0.06$ GeV~\cite{PDG}.

\begin{figure}[ht]
\centerline{\epsfysize=10cm\epsffile{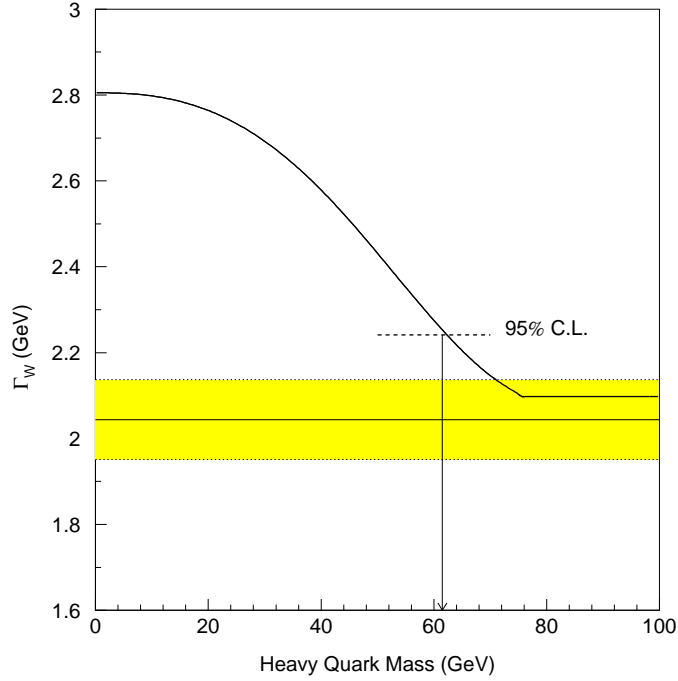}}
\caption{The width of the $W$ boson as a function of a new quark mass. 
Our measurement is shown as a one standard deviation band 
with the central value 
represented by the solid line.
The darker curve represents the prediction of the standard model 
as a function of quark mass.
The short dashed line indicates the upper limit at 95\% CL 
on the width of the $W$ boson from our data.}
 \label{fig:gammaw}
\end{figure}

We can use our result to probe
new possible decay modes of the $W$ boson,
such as decays into supersymmetric charginos
and neutralinos~\cite{susy} or heavy quarks~\cite{alvarez}.
Since our experimentally measured central value of 
$\Gamma(W)/\Gamma(W \rightarrow l\nu)$ (the inverse of the branching 
fraction) falls below the mean predicted by
the standard model,
we use the asymmetric method to calculate limits 
on new decay modes~\cite{PDG}. From our data, we derive a 95\% CL upper
limit of 171 MeV
on the width of unexpected decays of the $W$ boson.
If a new heavy quark exists, the limit for its mass is 
$m_{q'} > 61$ GeV at the $95\%$ CL (see Fig.~\ref{fig:gammaw}). 
Combining our result with other 
measurements~\cite{average} gives a weighted average of 
$ \Gamma(W) = 2.062 \pm 0.060$ GeV, 
and a 95\% CL upper limit of 111 MeV on unexpected decays.

Since the time that these results were first reported in a Letter
\cite{D0PRL}, the knowledge of the mass of the $W$ boson has improved 
substantially.  If we update the value used in Ref.~\cite{D0PRL} of
$M_W=80.23\pm0.18$ GeV to the current value of 
$M_W=80.39\pm0.06$ GeV\cite{mwmodern}, the following results are obtained:
\begin{equation} \sigma(W) \rm{B}(W\to e\nu) 
= 2.35 \pm 0.02 \pm 0.08 \pm 0.13\rm{nb}, \end{equation}
\begin{equation} B(W \to l \nu) = (11.03\pm0.52) \%. \end{equation}
\begin{equation} \Gamma(W) = 2.054\pm0.097 \mbox{ GeV} \end{equation}
All other numbers reported change by much less than their uncertainties.

\section{Conclusions}

D\O\ has measured the product of cross section
and the lepton branching fraction for $W$ and $Z$ boson production in the
electron and muon decay channels. We find
\begin{eqnarray*}
\sigma(W) \rm{B}(W\to e\nu) &=& 2.36 \pm 0.02 \pm 0.08 \pm 0.13\rm{nb}, \\
\sigma(W) \rm{B}(W\to \mu\nu) &=& 2.09 \pm 0.06 \pm 0.22 \pm 0.11\rm{nb}, \\
\sigma(Z) \rm{B}(Z\to ee) &=& 0.218 \pm 0.008 \pm 0.008 \pm 0.012\rm{nb},
\end{eqnarray*}
and
\begin{eqnarray*}
\sigma(Z) \rm{B}(Z\to \mu\mu) &=& 0.178 \pm 0.022 \pm 0.021 \pm 0.009\rm{nb}.
\end{eqnarray*}
Our values are in good agreement both with the ${\cal O}(\alpha_s^2)$
QCD predictions using recent pdf sets, and with other measurements.

Including theoretical calculations for $\sigma_W/\sigma_Z$ and B$(Z\to ll)$,
we measure 
\begin{equation} \rm{B}(W \to l \nu) = (11.02\pm0.52) \%. \end{equation}
Adding the standard model prediction
for $\Gamma(W \to l\nu)$, we find
\begin{equation} \Gamma(W) = 2.044\pm0.097 \mbox{ GeV}. \end{equation}
These results are in good agreement with the standard model,
and allow us to set a limit on any new decay modes of the $W$ boson.

\section{Acknowledgments}
%
%
We thank the staffs at Fermilab and collaborating institutions for their
contributions to this work, and acknowledge support from the 
Department of Energy and National Science Foundation (U.S.A.),  
Commissariat  \` a L'Energie Atomique (France), 
Ministry for Science and Technology and Ministry for Atomic 
   Energy (Russia),
CAPES and CNPq (Brazil),
Departments of Atomic Energy and Science and Education (India),
Colciencias (Colombia),
CONACyT (Mexico),
Ministry of Education and KOSEF (Korea),
and CONICET and UBACyT (Argentina).


\begin{references}
\bibitem{average} We recalculated $\Gamma(W)$ for the other
experiments~\cite{UA1,UA2,CDF1aR} by using their published $R$ values along
with the values of B$(Z\rightarrow ll)$ and
${\Gamma(W\rightarrow l\nu)}$ in this paper. For ${\sigma_W/ \sigma_Z}$ we
used our value for CDF, and
$3.26 \pm 0.09$~\cite{UA2} for the UA1 and UA2 experiments at 
lower~$\sqrt{s}$.
\end{references}
\end{document}